\documentclass[
reprint,
superscriptaddress,
 amsmath,amssymb,
 aps,
pra,
]{revtex4-2}

\usepackage{graphicx}
\usepackage{dcolumn}
\usepackage{bm}
\usepackage{xcolor}
\usepackage{hyperref}
\usepackage{subfigure}
\hypersetup{
    colorlinks,
    linkcolor={blue!80!black},
    citecolor={blue!80!black},
    urlcolor={blue!80!black}
}
\usepackage{braket}
\usepackage{extpfeil}
\usepackage{mathrsfs}

\usepackage{tikz}
\newcommand\parallelogram{
	\mathord{\text{
			\tikz[baseline]
			\draw (0,.1ex) -- (.8em,.1ex) -- (1em,1.6ex) -- (.2em,1.6ex) -- cycle;}}}

\begin{document}


\title{Stabilizing the Kerr arbitrary cat states and holonomic universal control}

\author{Ke-hui Yu}
\email{yukehui@stu.xjtu.edu.cn}
\affiliation{Institute of Theoretical Physics, School of Physics, Xi'an Jiaotong University, Xi'an 710049, China}
\author{Fan Zhu}
\affiliation{Institute of Theoretical Physics, School of Physics, Xi'an Jiaotong University, Xi'an 710049, China}
\author{Jiao-jiao Xue}
\affiliation{Engineering Research Center of X-ray Imaging and Detection, Universities of Shaanxi Province}
\affiliation{School of Physics and Electrical Engineering, Weinan Normal University}
\author{Hong-rong Li}
\email{hrli@mail.xjtu.edu.cn}
\affiliation{Institute of Theoretical Physics, School of Physics, Xi'an Jiaotong University, Xi'an 710049, China}
\date{\today}

\begin{abstract}
The interference-free double potential wells realized by the two-photon driving Kerr nonlinear resonator (KNR) can stabilize cat states and protect them from decoherence through a large energy gap. In this work, we use a parametrically driving KNR to propose a novel engineering Hamiltonian that can stabilize arbitrary cat states and independently manipulate the superposed coherent states to move arbitrarily in phase space. This greater degree of control allows us to make the two potential wells collide and merge, generating a collision state with many novel properties. Furthermore, the potential wells carrying quantum states move adiabatically in phase space produce quantum holonomy. We explore the quantum holonomy of collision states for the first time and propose a holonomy-free preparation method for arbitrary cat states. Additionally, we develop a universal holonomic quantum computing protocol utilizing the quantum holonomy of coherent and collision states, including single-qubit rotation gates and multi-qubit control gates. Finally, we propose an experimentally feasible physical realization in superconducting circuits to achieve the Hamiltonian described above. Our proposal provides a platform with greater control degrees of freedom, enabling more operations on bosonic modes and the exploration of intriguing physics.
\end{abstract}

\maketitle


\section{Introduction}
\begin{figure}
    \centering
    \includegraphics[width=\linewidth]{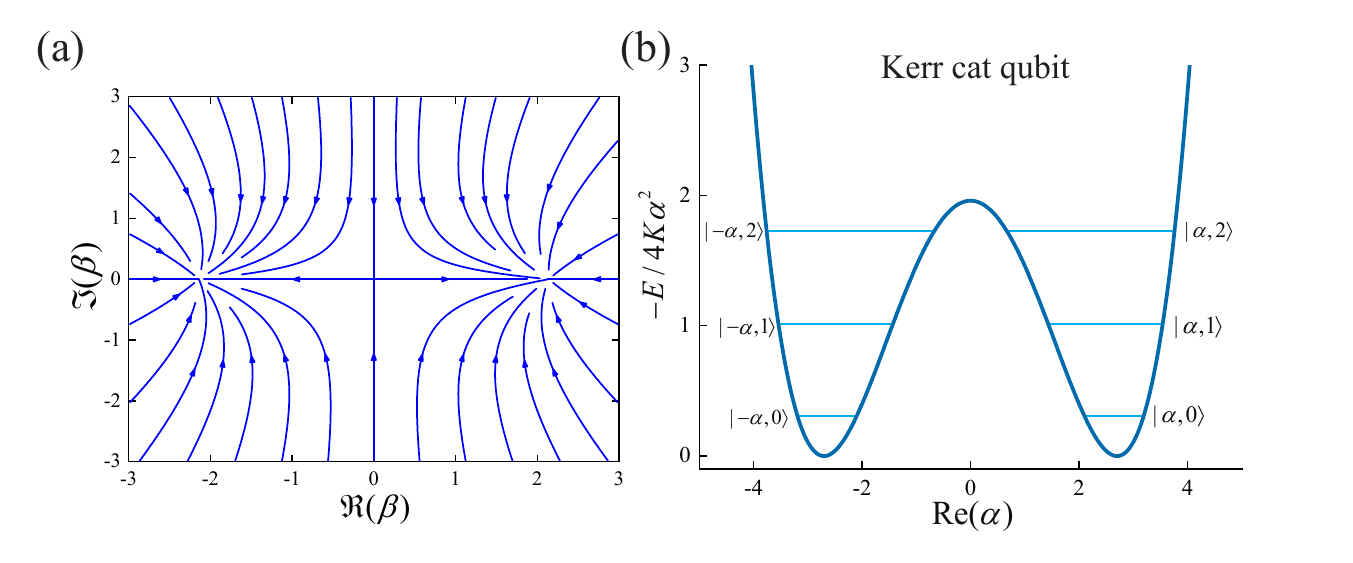}
    \caption{(a) Semi-classical vector field associated to two-photon dissipation. (b) Semi-classical double potential wells and eigenspectrum of two-photon driving KNR.}
    \label{fig:dissipative_kerr_potential}
\end{figure}

Quantum computing is much faster than classical computing in solving certain problems because of its inherent quantum superposition and coherence \cite{nielsen2010quantum,arute2019quantum,zhong2020quantum,wu2021strong}. However, it is not easy to achieve because quantum systems interact with uncontrollable freedom of environment to cause decoherence leading to undesired errors. To realize practical quantum computing, it is necessary to use quantum error correction (QEC) to protect qubits from errors \cite{roffe2019quantum}. Unlike discrete variable (DV) system that require multiple physical qubits for encoding \cite{steane1996multiple,fowler2012surface,kelly2015state}, using the redundancy from infinite energy levels of harmonic oscillators to encode information on a single physical qubit \cite{gottesman2001encoding,leghtas2013hardware,mirrahimi2014dynamically,michael2016new,puri2019stabilized}, that is, continuous variable (CV) system, can greatly reduce the overhead. Here, we focus on the cat code is feasible on near-term quantum devices, which encodes information on well-seperated cat states $\ket{\alpha}\pm \ket{-\alpha}$ with a larger $\alpha$ that can exponentially suppress dephasing error and also actively correct photon loss errors through parity detection \cite{leghtas2013hardware,puri2019stabilized,cochrane1999macroscopically}. Due to the strong photon-photon coupling and Kerr nonlinearity, superconducting circuits have the advantage to achieve cat states \cite{kirchmair2013observation}. There are many ways to prepare the cat state. The first is based on the qubit-cavity system, with qubit as the ancilla and cavity as the storage, the cat state can be prepared by qcMAP gate mapping the arbitrary qubit state into multi-component cat state with strong dispersive coupling \cite{leghtas2013deterministic} or echoed conditional displacement (ECD) gate with weak coupling \cite{eickbusch2022fast}. However, these schemes are very sensitive to photon loss, in order to avoid this shortcoming, the second scheme is the engineering two-photon dissipation by coupling a qubit to a high-quality cavity as storage and a low-quality resonator as dissipative bath, as shown in Fig.~\ref{fig:dissipative_kerr_potential}(a), this two-photon dissipation creates a vector potential field in phase space that tends toward two stable equilibria $\pm\alpha$, attracting a quantum states to slip into these two points and thus stabilizing to the cat state, which is called the dissipative cat state \cite{mirrahimi2014dynamically,leghtas2015confining}. The third scheme is to apply a two-photon driving to the Kerr nonlinear resonator (KNR). As shown in the Fig.~\ref{fig:dissipative_kerr_potential}(b), two-photon driving KNR generates a double well located at $\pm \alpha$ in phase space, and its ground eigenstate is cat state. When the detuning is zero, the quantum interference between potential wells is exponentially suppressed by $e^{-2|\alpha|^2}$. In addition, there is a large protective energy gap $4K|\alpha|^2$ between the ground state subspace and the higher energy levels, which can stabilize the cat state \cite{puri2017engineering,grimm2020stabilization}. Both the dissipative and Kerr cat state are protected by engineering Hamiltonians, in this paper, we focus on the Kerr cat state, and the main results are also applicable to dissipative cat state.
\par However, for the existing scheme, the cat state can only be trapped in $\pm\alpha$, and its two superposition components are artificially shackled. The pursuit of freedom is human nature, cat is no exception, a very natural idea is to remove the restriction of two superposition components, so that they can appear independently anywhere in phase space, we call it arbitrary cat state. 
There are certainly many interesting things to explore extending from the special to the general case, however, the existing two-photon driving KNR cannot implement arbitrary cat state. To overcome this challenge, in Sec.~\ref{sec:arbitrary cat states}, we propose the engineering Hamiltonian for stabilizing arbitrary cat states in parametric driving KNR, and analyze its fundamental properties such as eigenstates, energy spectrum structure, and stability under photon loss. Secondly, our proposed Hamiltonian has a greater degree of control freedom, which can make the potential wells move arbitrarily in phase space, or even collide and merge. So in Sec.~\ref{sec:control_geometric_phase}, we discuss the arbitrary control, because to keep the system in the ground state at every moment, we make the potential wells move adiabatically. For our degenerate systems, this naturally leads to geometric phases and quantum holonomy \cite{wilczek1984appearance,zhang2023geometric}. In our system, it is possible to realize the collision of potential wells, whose eigenstates we called collision states. We explored the quantum holonomy of adiabatic movement of collision states for the first time, which can be directly used to construct quantum gates of two-level systems. In addition, by studying the fundamental properties of collision state, we found that it can achieve a higher sensitivity for quantum metrology. However, before we go any further, there is till a very basic problem, how do we generate the arbitrary cat state? Based on the study of arbitrary control, we know that the adiabatic movement of quantum states will induce quantum holonomy, which is harmful to the preparation of arbitrary cat states. If the potential well is adiabatically separated from the origin to $\alpha_0, \alpha_1$ according to the traditional cat state preparation method \cite{puri2017engineering,xue2022fast}, uncontrollable holonomy will be brought about. In Sec.~\ref{sec:preparation}, we address this problem by proposing a holonomy-free adiabatic preparation path, which can achieve accurate target state preparation. Combine the idea in \cite{albert2016holonomic} and our proposed system, we construct universal holonomic gates in Sec.~\ref{sec:universal}. Among them, we propose $R_{y,z}$ gate and multi-qubit control gate for the first time. Finally, to realize this arbitrarily controlled Hamiltonian is not travail, in Sec.~\ref{sec:phys_realize}, we introduce asymmetric superconducting quantum interference device (a-SQUID) to provide third-order driving, and propose a physical circuit to realize our Hamiltonian, which can be implemented with the existing technologies. Our research provides a new platform with higher degree of control freedom to explore more operations of continuous variable systems and other interesting physics.

\section{Arbitrary cat states}
\label{sec:arbitrary cat states}
In quantum optics, the cat state is defined as the superposition of two coherent states with the same amplitude and opposite phases, $\ket{\rm CAT}=\mathcal{N}_\pm(\ket{\alpha}\pm\ket{-\alpha})$ \cite{cochrane1999macroscopically}, where $\mathcal{N}_\pm$ is the normalization coefficient. A natural progression is to generalize from this specific case. We introduce the arbitrary cat state (ACS), which is $\ket{\rm ACS}=\sum_j c_j \ket{\alpha_j}$ composed of coherent states with arbitrary amplitudes and phases in arbitrary proportions \cite{haroche2006exploring}. In this paper, we focus on the two-component ACS, which can be easily extended to an n-component ACS. the two-component ACS is defined as: 
\begin{equation}
\label{eq:ACS}
    \ket{\mathcal{C}} = c_0\ket{\alpha_0} + c_1\ket{\alpha_1},
\end{equation}
where it is a superposition of two arbitrary coherent states $\ket{\alpha_0}$, $\ket{\alpha_1}$ with arbitrary coefficients $c_0$, $c_1$. Compared with the symmetric cat state (SCS), ACS exhibits extended non-classical properties. For example, Prakash and Kumar found that amplitude-squared squeezing of $X\cos\theta + P\sin\theta$ has a minimum value of 0.3268 at $\alpha_0-\alpha_1=2.16\exp(\pm i(\pi/4) + i \theta/2)$ and $c_1/c_0=0.3\exp[1/2(\alpha_0\alpha_1^*-\alpha_0^*\alpha_1)]$ \cite{prakash2008amplitude}. Other experiments have also demonstrated the non-classical properties of such states and their applications in quantum information \cite{rivera2022strong,stammer2023quantum,martos2023metrological,lamprou2023nonlinear}. However, most of these are implemented in optical platforms and separation of $\alpha_0$ and $\alpha_1$ is very small. In this paer, we will focus on superconducting circuits and demonstrate for the first time how to achieve ACS and arbitrary control using Kerr nonlinear resonator.
\subsection{Engineering Hamiltonian in Kerr nonlinear resonator}
\begin{figure*}
    \centering
    \includegraphics[width=0.9\linewidth]{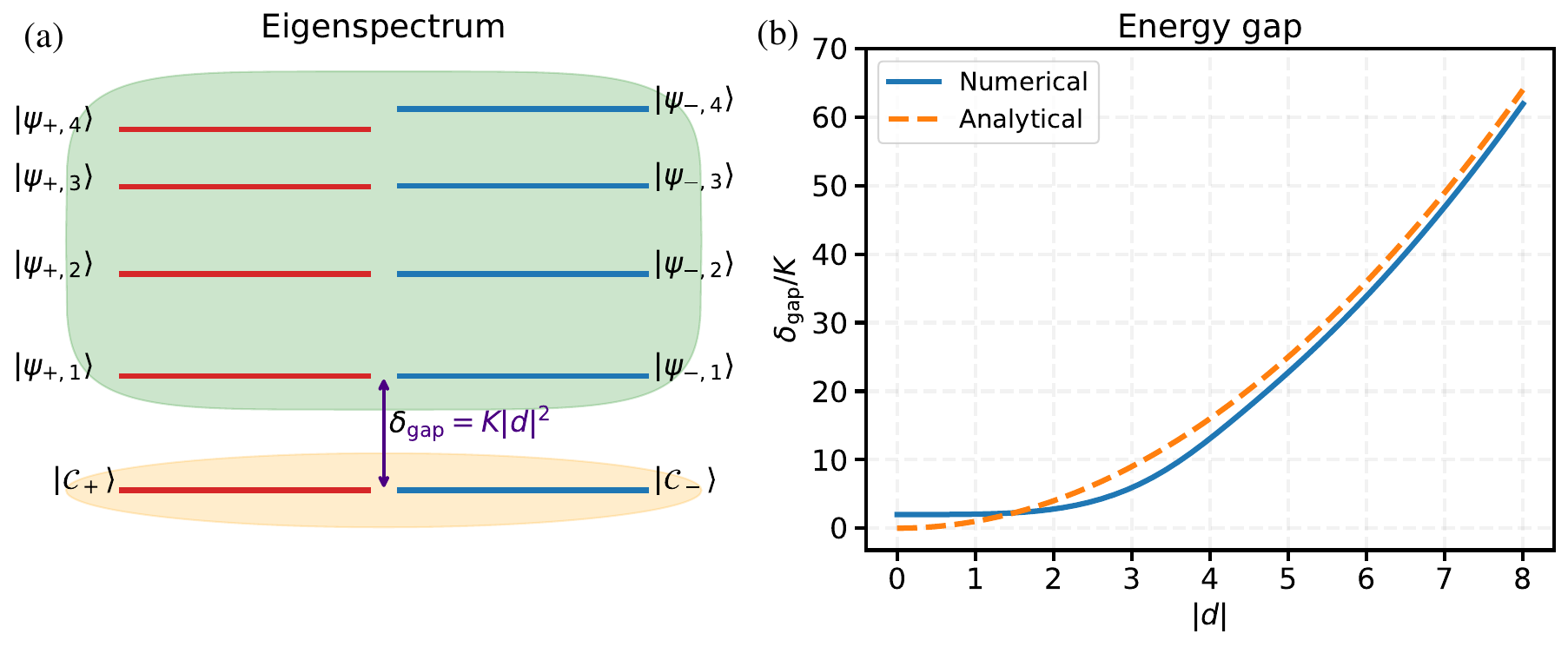}
    \caption{(a) Eigenspectrum obtained through exact numerical diagonalization of the Hamiltonian in Eq.~(\ref{eq:ham_KNR}) with $\alpha_0=4e^{i2\pi/3}$, $\alpha_1=3e^{-i\pi/4}$. Red lines represent plus states $\ket{\psi_{+,n}}$, blue lines represent minus states $\ket{\psi_{-,n}}$. The orange region denotes the ground state manifold, including ACS states $\ket{\mathcal{C}_\pm}$. The green region indicates higher energy levels, with an energy gap $\delta_{gap}=K|d|^2$ separating them from the ground state. (b) Energy gap $\delta_{gap}/K$ versus separation distance $|d|^2$. The blue solid line shows the numerical solution by exact diagnolization of Eq.~(\ref{eq:ham_KNR}), and the orange dashed line shows the analytical approximation. The solutions converge as $|d|$ increases and completely coincide as $|d|$ approaches infinity.}
    \label{fig:eigenspectrum}
\end{figure*}
Cat states can be implemented in many systems \cite{ourjoumtsev2007generation, leghtas2013deterministic, mirrahimi2014dynamically, puri2017engineering, puri2019stabilized, duan2019creating}. Recently, cat states based on two-photon driving in KNR have been widely studied \cite{goto2016universal, puri2017engineering, puri2019stabilized, wang2019quantum, puri2020bias, grimm2020stabilization, darmawan2021practical, putterman2022stabilizing, he2023fast, hajr2024high, aoki2024control}. This type of system produces a double potential well where the cat state is the ground eigenstate, separated by a large energy gap from higher energy levels. This energy gap helps protect and stabilize the cat state, known as the Kerr cat state, making it immune to resonator dephasing and enabling active error correction of photon loss through photon parity detection and feedback control \cite{leghtas2013hardware, mirrahimi2014dynamically, puri2019stabilized}. Therefore, we also focus on this system. However, currently, there is no established method for implementing ACS. In this paper, we address this challenge by introducing third-order driving and propose the engineering Hamiltonian for ACS implementation in parametric driving KNR for the first time. The Hamiltonian in the rotating frame with frequency $\omega_p$ is given by:
\begin{align}
\label{eq:ham_eigen}
\hat{H}&=K(\hat{a}^\dagger-\alpha_0^*)(\hat{a}^\dagger-\alpha_1^*)(\hat{a}-\alpha_0)(\hat{a}-\alpha_1) - K|\alpha_0\alpha_1|^2\\
\label{eq:ham_KNR}
&=|\eta|^2\hat{a}^\dagger\hat{a}+K\hat{a}^{\dagger2}\hat{a}^2+[(\beta\eta^*\hat{a}^\dagger+\beta\hat{a}^{\dagger 2}+\eta\hat{a}^{\dagger2}\hat{a})+h.c.],
\end{align}
where $K$ is the Kerr nonlinearity, $\beta=K\alpha_0\alpha_1$ and $\eta=-K(\alpha_0+\alpha_1)$ denote the second and third-order driving strengths respectively. In addition, to fully engineer the Hamiltonian, it is essential to include the parameter-mathched detuning $|\eta|^2$ and first-order driving with strength $\beta \eta^*$. As evident from Eq.~(\ref{eq:ham_eigen}), it is obvious that $\ket{\alpha_0}$ and $\ket{\alpha_1}$ are the eigenstates of the system, and consequently, arbitrary cat states also serve as eigenstates. Hence, the Hamiltonian enables ACS implementation, offering arbitrarily control over $\alpha_0$ and $\alpha_1$ through tailored $\beta$ and $\eta$ values for driving the KNR. This heightened control, explored in detail in Sec.~\ref{sec:control_geometric_phase}, grants us expanded flexibility in manipulation. However, to realize the Hamiltonian in Eq.~(\ref{eq:ham_KNR}) is not travail, unlike the typical two-photon driving KNR associated with the SCS, it necessitates additional single-photon driving, detuning, and third-order driving terms. Control over detuning is achieved by tuning the frequency of the rotating frame, and the driving frequencies can be set to $\omega_1=\omega_2/2=\omega_3=\omega_p$ to realize all driving terms, where $\omega_n$ represent the driving frequency of the $n$th-order driving. In Sec.~\ref{sec:phys_realize}, we introduce an a-SQUID to facilitate third-order driving and propose a comprehensive superconducting circuit to realize the Hamiltonian in Eq.~(\ref{eq:ham_KNR}), which is achievable with existing technologies.

\subsection{Energy spectrum and eigenstates}
The coherent state can be expressed as a displacement operation on the vacuum state $\mathcal{D}(\alpha)\ket{0}$, where the displacement operator is defined as $\mathcal{D}(\alpha) = \exp(\alpha \hat{a}^\dagger - \alpha^* \hat{a})$. Furthermore, we introduce the general shifted Fock basis $\ket{\psi_{\pm,n}}=\mathcal{N}_{\pm,n}[\mathcal{D}(\alpha_0)\ket{n}\pm (-1)^n \mathcal{D}(\alpha_1)]\ket{n}$, where $\ket{n}$ is the Fock state with $n$ photons, $\mathcal{N}_{\pm,n}$ is the normalization factor. This set forms a complete basis and approximates the eigenstates of the Hamiltonian in Eq.~(\ref{eq:ham_KNR}), becoming exact eigenstates as the separation $|d|=|\alpha_0-\alpha_1|$ of ACS approaching infinity. Through exact numerical diagonalization of the Hamiltonian in Eq.~(\ref{eq:ham_KNR}) with arbitrarily setting $\alpha_0=4e^{i2\pi/3}$, $\alpha_1=3e^{-i\pi/4}$, we obtain the eigenspectrum shown in  Fig.~\ref{fig:eigenspectrum}(a). In this figure, the red lines represent the plus states $\ket{\psi_{+,n}}$, the blue lines represent the minus states $\ket{\psi_{-,n}}$. Moreover, the orange region denotes the ground state manifold, where numerical results show that the ground state is strictly degenerate. Thus, the lowest energy states in the shifted Fock basis with $n = 0$, i.e., ACS $\ket{\mathcal{C}_\pm}=\mathcal{N}_{\pm}( \ket{\alpha_0} \pm \ket{\alpha_1})$, with $\mathcal{N}_{\pm}$ defined in Appdendix~\ref{appdix:orth_norm}, are the exact ground states for finite separation $|d|$, so $\ket{\alpha_0}, \ket{\alpha_1}$ are also exact ground states. The green region represents the rest of higher energy levels manifold, with a large energy gap $\delta_{\rm gap}$ separating it from the ground states, thus protecting the system and stabilizing it in the ground subspace. To illustrate it more intuitively, we apply the displacement transformation to Eq.~(\ref{eq:ham_KNR}) to obtain the displaced Hamiltonian
\begin{align}
    \label{eq:ham_displaced_start}
    \hat{H}_\alpha^\prime &= \mathcal{D}(\alpha)^\dagger \hat{H} \mathcal{D}(\alpha)\\ 
    &= (|\eta|^2 + 4K|\alpha|^2 + 2 \eta^* \alpha + 2\eta \alpha^*)\hat{a}^\dagger\hat{a} + K\hat{a}^{\dagger 2}\hat{a}^2 \\
    \label{eq:ham_displace_single_photon}
    \begin{split}
        &+(\alpha |\eta|^2 + 2K\alpha_i^2\alpha^* + \beta \eta^* + 2\beta \alpha^* + 2\eta|\alpha|^2\\&+\eta^*\alpha^2)\hat{a}^\dagger + h.c.
    \end{split}\\
    &+(K\alpha^2 + \beta + \eta \alpha)\hat{a}^{\dagger 2} + h.c. \\
    \label{eq:ham_displaced_end}
    &+ (2K\alpha + \eta)\hat{a}^{\dagger 2}\hat{a} + h.c.
\end{align}
We drop the constant term because it just represents a shift in the absolute value of the energy spectrum and does not affect the energy levels spacing. Setting $\alpha$ to $\alpha_0$ and $\alpha_1$ respectively, assuming a larger separation distance $|d|$, we transform the Hamiltonian of Eq.~(\ref{eq:ham_KNR}) into shifted Fock basis to get $\hat{H}^\prime=1/2(\hat{H}_{\alpha_0}^\prime + \hat{H}_{\alpha_1}^\prime)$. The ground state is vaccum state $\ket{0}$ and the energy is $\bra{0}\hat{H}^\prime \ket{0}=0$. The approximate first excited state is Fock state $\ket{1}$ which energy $\bra{1}\hat{H}^\prime \ket{1}=K|\alpha_0-\alpha_1|^2=K|d|^2$. Therefore, the energy gap between the ground state subspace and the higher energy levels is $\delta_{\rm gap}=K|d|^2$. Fig.~\ref{fig:eigenspectrum}(b) shows the change of the energy gap with the separation distance $|d|$. The exact numerical solution and the approximate analytical solution gradually converge as $|d|$ increases, and will completely coincide as $|d|$ approaches infinity. The larger the separation distance $|d|$, the better the protection effect of the energy gap on the ground state subspace. Therefore, using the Hamiltonian in Eq.~(\ref{eq:ham_KNR}) we can stabilize the system in the ACS.

\subsection{Stabilization with photon loss}
\label{sec:stablize_loss}
In the resonator, photon loss is the predominant error channel. Taking it into account, the master equation takes the form $\dot{\rho}=-i[\hat{H}_0,\rho]+\sqrt{\kappa}D[\hat{a}]$, with the non-Hermitian effective Hamiltonian given by $\hat{H}_{\rm eff}=\hat{H}_0-i\kappa \hat{a}^\dagger \hat{a}/2$, where the $\hat{H}_0$ is our engineering Hamiltonian from Eq.~(\ref{eq:ham_KNR}). Performing a displacement transformation, the Hamiltonian $\hat{H}_0$ transforms to $\hat{H}_{\alpha}^\prime$ of Eq.~(\ref{eq:ham_displaced_start}), and the term $-i\kappa \hat{a}^\dagger \hat{a}/2$ transforms to $-(i\kappa \alpha \hat{a}^\dagger/2 + h.c.) - i\kappa \hat{a}^\dagger \hat{a}/2$, this results in the displaced effective Hamiltonian:
\begin{align}
    \label{eq:ham_effective_displaced}
    \hat{H}_{\rm eff}^\prime=\hat{H}_{\alpha}^\prime-i\kappa \hat{a}^\dagger \hat{a}/2-(i\kappa \alpha \hat{a}^\dagger/2 + h.c.).
\end{align}
For the system to remain stable after the displacement transformation, i.e., no more displacement occurs, the generator of the displacement operator--the single-photon term—must be canceled. Actually, since $\ket{\alpha_0}$ and $\ket{\alpha_1}$ are eigenstates of Eq.~(\ref{eq:ham_KNR}), the single photon terms of Eq.~(\ref{eq:ham_displace_single_photon}) in displaced Hamiltonian $\hat{H}_{\alpha_0}^\prime$ and $\hat{H}_{\alpha_1}^\prime$ are both zero, indicating stability at $\alpha_0$ and $\alpha_1$. To stabilize the system with photon loss, we need to find $\alpha$ such that the single-photon term in $\hat{H}_{\rm eff}^\prime$ is zero. This involves solving the following equation:
\begin{align}
\label{eq:steady_loss}
\begin{split}
    &\alpha |\eta|^2 + 2K\alpha_i^2\alpha^* + \beta \eta^* + 2\beta \alpha^* + 2\eta|\alpha|^2+\eta^*\alpha^2 - i\frac{\kappa}{2}\alpha\\&=0.
\end{split}
\end{align}
\begin{figure}
    \centering
    \includegraphics[width=\linewidth]{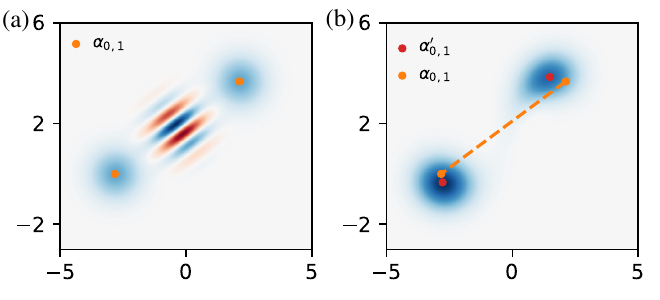}
    \caption{Stabilization with photon loss. (a) Wigner function of the final state without photon loss, where the system remains unchanged and stabilizes in initial cat state with $\alpha_0=3e^{i\pi/3}, \alpha_1=-2$ (orange dots). (b) Wigner function of the final state with a photon loss rate $\kappa = 5K$ at $T = 4/\kappa$. Red dots indicate the expected steady-state positions $\alpha_0^\prime$ and $\alpha_1^\prime$ solved from Eq.~(\ref{eq:steady_loss}), showing rotation, displacement, and a reduced separation distance compared to the original states $\alpha_0$ and $\alpha_1$ (orange dots), but still maintaining a significant separation.}
    \label{fig:steady_loss}
\end{figure}
This equation has solutions, which can be found numerically to determine the steady state. We verify this with a case $\alpha_0=3e^{i\pi/3}$, $\alpha_1=-2$. We simulate the time evolution of the master equation with and without photon loss and plot the Wigner function of the final state. Fig.~\ref{fig:steady_loss}(a) shows the case without loss, where the initial quantum state remains unchanged after the evolution. Fig.~\ref{fig:steady_loss}(b) depicts the final state with a loss rate $\kappa=5K$ at $T=4/\kappa$. The steady-state positions $\alpha_0^\prime$ and $\alpha_1^\prime$ (red dots), obtained by solving the Eq.~(\ref{eq:steady_loss}), align perfectly with the master equation simulation. Compared with the original state $\alpha_0, \alpha_1$ (orange points), the steady state under loss shows rotation, displacement, and reduced separation distance but still maintains a significant separation. These results confirm that the system can stabilize in the cat states $\mathcal{N}_{\pm}(\ket{\alpha_0^\prime}\pm \ket{\ket{\alpha_1^\prime}})$ manifold even with photon loss.
For a small amount of resonator dephasing noise $\sqrt{\kappa_{\varphi}}D[ \hat{a}^\dagger \hat{a}]$, we project the noise operator to the ACS subspace:
\begin{align}
\begin{split}
    \hat{P}_{\mathcal{C}} \hat{a}^\dagger \hat{a} \hat{P}_{\mathcal{C}}&=(|\alpha_0|^2+|\alpha_1|^2)\hat{I} + 2|\alpha_0 \alpha_1|\cos(\gamma)e^{-\frac{1}{2}|d|^2}\hat{Z}\\
    &+(|\alpha_0|^2-|\alpha_1|^2)\hat{X} + 2|\alpha_0 \alpha_1|\sin(\gamma)e^{-\frac{1}{2}|d|^2}\hat{Y},
\end{split}
\end{align}
where the projection operator is $\hat{P}_{\mathcal{C}}=\ket{\mathcal{C}_+}\bra{\mathcal{C}_+}+\ket{\mathcal{C}_-}\bra{\mathcal{C}_-}$, and the factor $\gamma = \rm{Im}[\alpha_0 \alpha_1^*]=|\alpha_0\alpha_1| \sin(\phi_0-\phi_1)$ with the phase $\phi_i$ of $\alpha_i$, see Appdendix~\ref{appdix:orth_norm} for more detail. To facilitate the analysis of the noise, we consider the $\ket{\mathcal{C}_\pm}$ as a qubit, since they are strictly orthogonal, we observe that the phase flip error in this system is exponentially suppressed by the separation distance $|d|$. The bit flip error, however, is proportional to $|\alpha_0|^2-|\alpha_1|^2$. This indicates that the asymmetry of the ACS enhances the bit flip noise, while the traditional SCS, i.e., $\alpha_0=-\alpha_1$, is completely immune to resonator dephasing. Despite this, the asymmetry of ACS does not introduce unbiased noise, the main noise channel remains bit-flip same as the traditional cat qubit, can be corrected by concatenated QEC codes \cite{puri2020bias}. Actually, ACS does not offer advantages as a qubit, so we introduce it does not imply that we use it to encode logical qubit. It represents a tradeoff: while increasing a bit of biased noise, it provides greater control freedom by decoupling $\alpha_0$ and $\alpha_1$, thus enabling more versatile applications of cat states. In Sec.~\ref{sec:universal}, we leverage this control freedom to achieve holonomic universal quantum computing with cat qubits. In this protocol, ACS serves only as an intermediate quantum state, and we avoid potential negative effects from asymmetry through careful design of the evolutionary path.

\section{Arbitrary control and quantum holonomy}
\label{sec:control_geometric_phase}
\begin{figure*}[ht]
    \centering
    \includegraphics[width=0.9\linewidth]{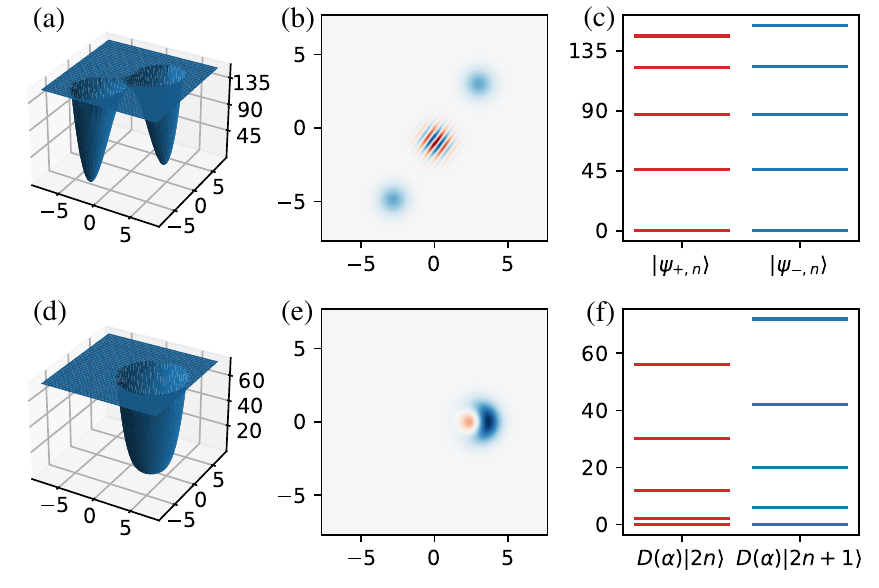}
    \caption{The semiclassical metapotential surface of (a) ACS with $\alpha_0=4e^{-i 2\pi/3}$, $\alpha_1=3e^{i\pi/4}$ and (d) collision state with $\alpha=2$. The Wigner functions of ACS and collision state are depicted in (b) and (e), respectively, while (c) and (f) display the corresponding eigenspectrum of the ACS and collision state Hamiltonians.}
    \label{fig:potential_wigner}
\end{figure*}
As can be seen from the Eq.~(\ref{eq:ham_eigen}), different from the SCS which have the restriction $\alpha_0=-\alpha_1$, our system decouples $\alpha_0$ and $\alpha_1$, giving them independent control degrees of freedom. In this section, we will provide a more vivid analysis by introducing the semi-classical metapotential surface in Section \ref{sec:metapotential}. This will help to visualize how our Hamiltonian, with its double-well structure, can stabilize the quantum state at arbitrary positions. Next, in Sec.~\ref{sec:collision_states}, we investigated a phenomenon that is unique to our system: the ability of arbitrary control allows us to make the double well collide and merge at any point, in which case we define the eigenstate of the system as the collision states, and we explore its fundamental properties in detail. Building on the control freedom of our system, it serves as an exceptional platform for studying the intriguing phenomenon of quantum holonomy, which arises from the adiabatic evolution of the system's wavefunction in parameter space. Quantum holonomy encompasses both Abelian and non-Abelian forms. The Abelian quantum holonomy, known as the Berry phase, was introduced by Michael Berry in 1984 \cite{berry1984quantal}. It involves the geometric phase acquired by a system’s wavefunction after undergoing a cyclic adiabatic process. Wilczek and Zee extended this concept to degenerate states, where adiabatic evolution results in non-Abelian quantum holonomy, represented as a unitary matrix \cite{wilczek1984appearance}. Due to its global features, quantum holonomy is of significant interest in the field of quantum computation, offering potential robustness against certain local errors. In Sec.~\ref{sec:holonomy}, we investigate the quantum holonmy by making the potential wells move adiabatically. Specifically, we demonstrate the geometric phase of the coherent state in the case of potential well separation, which corresponds to Abelian quantum holonomy. Furthermore, for the case of potential well colliding, we reveal for the first time the non-Abelian quantum holonomy of collision states. These findings have led to the development of more valuable applications, including precise arbitrary cat state preparation (Sec.~\ref{sec:preparation}) and universal holonomic quantum computing (Sec.~\ref{sec:universal}). Moreover, the greater control freedom in our system provides a new platform for exploring more operations and uncovering intriguing physics.

\subsection{Semi-classical metapotential surface}
\label{sec:metapotential}
To describe arbitrary control more vividly, we introduce the semi-classical metapotential surface \cite{venkatraman2022quantum}, from another perspective to visualize how the Hamiltonian stabilizes and controls the cat state. Applying the invertible Wigner transformation $\mathscr{D}[\hat{a}]=a=(x+ip)/\sqrt{2}$, we obtain the phase space formulation in classical limit
\begin{align}
\label{eq:ham_cl}
\begin{split}
    H&=|\eta|^2 \frac{x^2+p^2}{2} + \frac{(x^2+p^2)^2}{4} + \left[\beta \eta^* \frac{x-ip}{\sqrt{2}} + c.c.\right]\\
    &+\left[\beta\frac{(x+ip)^2}{2} +c.c.\right] + \left[\eta\frac{(x-ip)(x^2+p^2)}{2\sqrt{2}}+c.c. \right].
\end{split}
\end{align}
In the case of $\alpha_0 \neq \alpha_1$, corresponding to an arbitrary cat state, we can plot the metapotential forms double potential wells in Fig.~\ref{fig:potential_wigner}(a), with the cat state occupying the lowest energy bound state, as depicted in Fig.~\ref{fig:potential_wigner}(b). The energy spectrum of the corresponding Hamiltonian is presented in Fig.~\ref{fig:potential_wigner}(c), with the metapotential in Fig.\ref{fig:potential_wigner}(a) truncated to the 10th energy level of the spectrum. Notably, within the region of the double wells where the separation is evident, particularly across the first 8 energy levels, each pair of plus and minus states exhibits quasi-degeneracy (recall the exact degeneracy of the ground state). Referring to Eq.~(S3) in \cite{venkatraman2022quantum}, we have a similar Hamiltonian form of Eq.~(\ref{eq:ham_eigen}), which precludes tunneling between the two potential wells and thus inhibits the exchange of $\alpha_0$ and $\alpha_1$. In addition, the high potential barrier (corresponding to a large separation distance) ensures that cat state is difficult to jump to a higher energy level, guaranteeing its stability in the ground state. In addition, from a more intuitive perspective, the coherent or cat states is trapped at the bottom of the potential wells. When we move the well, the coherent state is pushed by the wall of the potential well. However, as shown in Fig.~\ref{fig:potential_wigner}(a), our wells have an open shape with a slope. If we move the wells too fast, the coherent state will slide up the slope and get excited to a higher energy level. To prevent this, adiabatic movement is required, allowing sufficient relaxation time for the coherent state to slip down to the ground state. Adiabatic control enables us to move the coherent state arbitrarily in phase space, achieving a greater degree of control freedom.

\subsection{Collision states}
\label{sec:collision_states}
As $\alpha_0$ and $\alpha_1$ can be controlled arbitrarily, we have the capability to gradually bring the two potential wells closer together until they collide or even merge. When $\alpha_0$ and $\alpha_1$ are sufficiently close, we can consider them to be colliding at a position $\alpha$ with a separation of $\Delta \alpha$. This situation characterizes the near-collision state as
\begin{align}
    \label{eq:near_col}
    \ket{\Psi}=\mathcal{N}_\alpha (\ket{\alpha+\Delta \alpha}-\ket{\alpha})
\end{align}
where $\Delta \alpha = |\Delta \alpha| e^{i\Delta \theta}$ represents a tiny separation. Here we pick up a minus state because its normalization coefficient $\mathcal{N}_\alpha$, is non-trivial, diverges to infinity as $|\Delta \alpha| \rightarrow 0$, thereby better representing the state in proximity to collision. Firstly, we review Othman and Yevick's theoretical work to elucidate the fundamental properties \cite{othman2018quantum}, such state in Eq.~(\ref{eq:near_col}) can be derived as
\begin{align}
    \label{eq:near_col_deriv}
    \ket{\Psi}=\frac{e^{i\delta \theta} \frac{\partial \ket{\alpha}}{\partial |\alpha|} + i |\alpha| \sin(\delta \theta)\ket{\alpha}}{\sqrt{1+|\alpha|^2\sin(\delta \theta)^2}},
\end{align}
where $\delta \theta=\Delta \theta-\theta$ and $\alpha=|\alpha|e^{i\theta}$. In the above equation, the near-collision state emerges as a superposition of the derivative state $\partial \ket{\alpha}/\partial |\alpha|$ and the coherent state $\ket{\alpha}$. The derivative state can be expanded as $\partial \ket{\alpha}/\partial |\alpha| = \left( \alpha/|\alpha|\hat{a}^\dagger - |\alpha|\right)\ket{\alpha}$. With $\hat{a}^\dagger \ket{\alpha}=\ket{\alpha,1}+\alpha^*\ket{\alpha}$, this derivative state $\partial \ket{\alpha}/\partial |\alpha|=e^{i\theta}\ket{\alpha,1}$ represents a displaced Fock 1 state with a global phase, commonly referred to as the Agarwal state \cite{agarwal1991nonclassical}. Consequently, we obtain
\begin{align}
    \ket{\Psi}=\frac{e^{i\Delta \theta}\ket{\alpha,1}+i|\alpha|\sin \delta \theta \ket{\alpha}}{\sqrt{1+|\alpha|^2\sin(\delta \theta)^2}}.
\end{align}
In the scenario of full collision $\delta \theta= 0$, $\ket{\Psi}=\ket{\alpha,1}$ a pure Agarwal state up to the global phase. Conversely, when $\delta \theta=\pm\pi/2$, $\ket{\Psi}=(\exp(i\theta)\ket{\alpha,1}+|\alpha|\ket{\alpha})/\sqrt{1+|\alpha|^2}$ denotes the maximum superposition state. Both scenarios reduce to Fock 1 at $\alpha=0$.
\par
\begin{figure}
    \centering
    \includegraphics[width=0.9\linewidth]{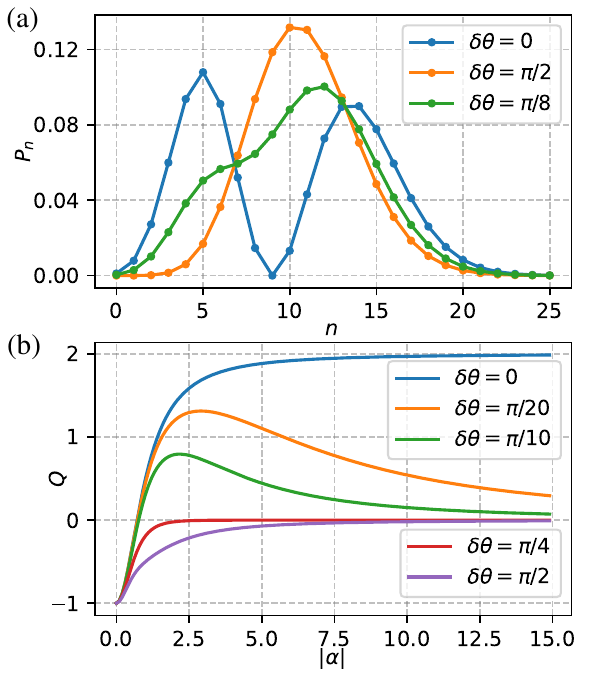}
    \caption{(a) Photon number probability distribution under different $\delta \theta$ with $\alpha=3e^{i2\pi/3}$. (b) The Mandel Q parameter varies with $|\alpha|$ for different $\delta \theta$.}
    \label{fig:near_col}
\end{figure}

 From the insights gained above, it becomes intuitive that the near-collision state will exhibit a variety of non-classical properties. We begin by examining its statistical characteristics. Fig.~\ref{fig:near_col}(a) shows the  of photon probability distribution $P_n=|\braket{n|\Psi}|^2$ with $\alpha=3e^{i2\pi/3}$. Notably, when $\delta \theta=0$, double peaks emerge, with the probability distribution reaching 0 at the central position $n=|\alpha|^2$, indicative of a pronounced quantum interference effect. As $\delta \theta$ increases, the distribution tends towards a unimodal form, resembling an near Poisson distribution when $\delta \theta=\pi/2$. The larger the $\alpha$, the closer it is to the coherent state, exhibiting near-classical properties. In addition, the average photon number is given by $\braket{\hat{a}^\dagger \hat{a}}=|\alpha|^2 + M$ with $M=(1+2|\alpha|^2\sin^2\delta \theta) / (1+|\alpha|^2\sin^2\delta \theta)$, and the photon fluctuation is expressed as $\braket{\Delta \hat{a}^\dagger \hat{a}}=\sqrt{\braket{\hat{a}^\dagger \hat{a}}(5-2M)+M(M-4)}$. The maximum photon fluctuation $\sqrt{3}|\alpha|$ occurs when $\delta \theta=0$ with the average photon number at its minimum value $|\alpha|^2 + 1$. Conversely, the minimum photon fluctuation arises when $\delta=\pi/2$ coinciding with the maximum photon number. For $|\alpha|\gg 1$, the fluctuation tends towards $\sqrt{|\alpha|^2-2}$ and the average photon number towards $|\alpha|^2+2$. Furthermore, let us delve into its quantum properties. the quantum uncertainty of near collision state is bounded by $1/16\leq \braket{(\Delta \hat{X})^2}\braket{(\Delta \hat{P})^2}\leq 9/16$. When $\delta \theta$ or $|\alpha|$ is zero, the maximum value $9/16$ is attained. Conversely, when $\delta \theta=\Delta \theta=\pi/2$ and $|\alpha|\gg 1$, the minimum value $1/16$ can be approached, it is worth noting that coherent states occupy the minimum uncertainty. Finally, we focus on measuring the nonclassicality. The Mandel Q parameter \cite{mandel1979sub}, which gauges the deviation of the photon number distribution from Poissonian statistics, is employed for the near-collision state as
\begin{align}
    Q=\frac{2|\alpha|^2(2-M)-M^2}{|\alpha|^2 + M}
\end{align}
We illustrate the relationship between $Q$ and $|\alpha|$ for various $\delta \theta$ in Fig.~\ref{fig:near_col}(b). When $|\alpha|$ approaches zero, the near-collision state resembles the Fock 1 state, characterized by $Q = -1$, indicating a notable anti-bunching characteristic.
For large values of $|\alpha|$, within the range $\pi/4<\delta \theta <3\pi/4$, $Q<0$ signifies an anti-bunching feature, while $\delta \theta<pi/4$ demonstrates a bunching feature. These characteristics reflect the non-classical properties of the near-collision state, which gradually approaches $Q=0$ with increasing $|\alpha|$, moving closer to the classical realm. For the full collision state with $\delta \theta=0$, $Q$ gradually approaches 2 as $|\alpha|$ increases, still displaying strong non-classical properties.
\par Based on these properties, Martos et al. discovered that the near-collision state exhibits superior metrological capabilities compared to traditional cat states. The Quantum Fisher Information (QFI) for phase estimation, with $\Delta \alpha=0.5$, exceeds that of cat states with the same $\alpha$ by more than 1.88 times. Additionally, unlike cat states, which become increasingly sensitive to photon number loss as $|\alpha|$ increases, the near-collision state exhibits greater robustness to photon number errors. Therefore, near-collision states offer comprehensive advantages in quantum metrology \cite{martos2023metrological}. However, in previous research, they generated the near-collision state using high harmonic generation (HHG) on an optical platform, which cannot achieve complete collision \cite{lewenstein2021generation, rivera2022strong}. What we aim for is a complete collision state with $\Delta \alpha=0$, expected to yield superior quantum metrology. Besides the aforementioned minus near-collision state, resulting in $\ket{\alpha,1}$, there exists a plus state, trivially corresponding to the coherent state $\ket{\alpha}$ upon collision. In Fig.~\ref{fig:potential_wigner}(d), the merging of the double potential wells into a wider single well is illustrated with $\alpha_0=\alpha_1=2$. In this case, the states $\ket{\alpha,n}$ represent the eigenstates of the Hamiltonian described in  Eq.~\ref{eq:ham_eigen}. Fig.~\ref{fig:potential_wigner}(f) displays the numerically diagonalized energy spectrum of the Hamiltonian, featuring a strictly degenerate ground state. Consequently, $\ket{\alpha}$ and $\ket{\alpha,1}$ emerge as a pair of degenerate ground states within the potential well, with a energy gap of 2K to a higher energy manifold. We can define plus and minus collision states as follows
\begin{align}
    \label{eq:col_state}
    \ket{\mathscr{C}_\pm} = \frac{1}{\sqrt{2}}(\ket{\alpha} \pm \ket{\alpha,1}),
\end{align}
the Wigner function of $\ket{\mathscr{C}_+}$ is depicted in Fig.~\ref{fig:potential_wigner}(e), exhibiting non-classical properties. It can be expected that, in addition to the fundamental quantum properties and advantages in quantum metrology mentioned above, these degenerate collision states offer a promising platform for exploring quantum information technology.
\subsection{Quantum holonomy}
\label{sec:holonomy}
\begin{figure}
    \centering
    \includegraphics[width=0.7\linewidth]{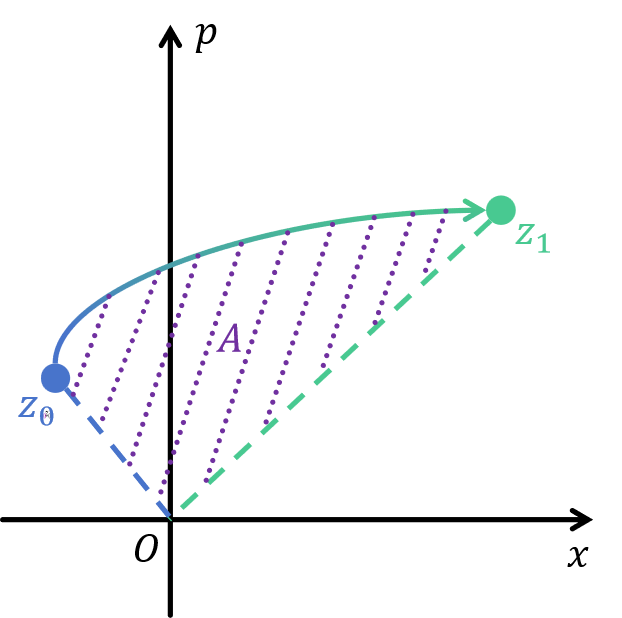}
    \caption{The trajectory of coherent state. The state moves from the $\ket{Z_0}$ (blue dot) to $\ket{Z_1}$ (green dot) along the trajectory (blue-green gradient line). The trajectory with $OZ_0$ (blue dashed line) and $OZ_1$ (green dashed line) encloses area $A$ (purple shaded area), resulting in a geometric phase of $-2A$.}
    \label{fig:geo_phase_coh}
\end{figure}

\begin{figure}
    \centering
    \includegraphics[width=\linewidth]{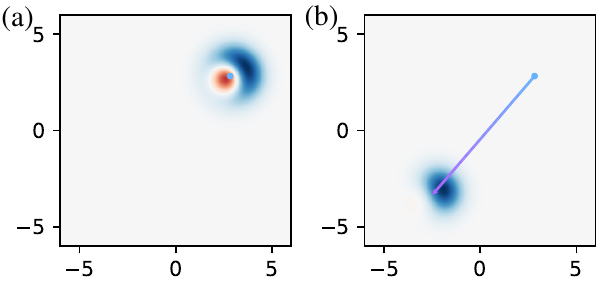}
    \caption{The Wigner function of the initial collision state (a), starting from 2+2i (blue dot) and evolving adiabatically to the final state (b) positioned at -1.7-2.3i (purple dot) along the blue-purple gradient path.}
    \label{fig:col_move}
\end{figure}

As discussed in Sec.~\ref{sec:metapotential}, we can control $\alpha_0$ and $\alpha_1$ allowing the potential wells carrying coherent states to move adiabatically in phase space. This process results in quantum holonomy \cite{zhang2023geometric, wilczek1984appearance}
\begin{align}
\label{eq:holonomy_def}
    U = \mathcal{P}\exp\left(i\int_{Z(0)}^{Z(T)} \Gamma(\mathbf{Z}^\prime) d \mathbf{Z}^\prime\right),
\end{align}
where $\mathcal{P}$ denotes path ordering on the parametric space $\mathbf{Z}$, and $\Gamma_{mn}(\mathbf{Z}^\prime)=i\bra{\psi_m}\nabla \ket{\psi_n}$ serve as a generator of unitary matrix for the n-dimensional degenerate space $\{ \psi_n \}$. In the scenario where $\alpha_0$ and $\alpha_1$ are widely separated, the two potential wells do not interfere with each other \cite{venkatraman2022quantum}. As depicted in Fig.~\ref{fig:dissipative_kerr_potential}(b), although the double wells exhibit degenerate ground states, each individual well's ground state is one-dimensional. Consequently, the adiabatic evolution of the coherent states in each potential well yields a geometric phase (Abelian quantum holonomy), which can be expressed as \cite{chaturvedi1987berry}
\begin{equation}
\label{eq:geo_phase_coh}
\varphi = \int_{\vec{l}} PdX - XdP = -2A, 
\end{equation}
where $X$ and $P$ represent quadratures of phase space. We illustrate this process in Fig.~\ref{fig:geo_phase_coh}, the coherent state starts at the blue dot $Z_0$ and moves to green dot $Z_1$ along the path delineated by the blue-green gradient line. The resulting geometric phase is negative twice the area $A$, which is represented by the purple shaded region enclosed by $OZ_0$, $OZ_1$ and the path. Where the sign of the area $A$ is determined by the path direction: counterclockwise is positive, while clockwise is negative.
Next, Let's consider another scenario where the separation distance of the potential wells is small or even colliding, resulting in a distinctly different situation. For clarity, we focus solely on the case of full collision, where $\alpha_0=\alpha_1=\alpha$. As described in Sec.~\ref{sec:collision_states}, the double wells colliding into a wider single well, with $\ket{\alpha}$ and $\ket{\alpha,1}$ being the degenerate ground states. Consequently, the evolution in this scenario yields a unitary matrix (non-Abelian quantum holonomy) in the degenerate subspace. Follow the definition of Eq.~(\ref{eq:holonomy_def}), the quantum holonomy of collision states is
\begin{align}
\label{eq:col_holonomy}
    U=\mathcal{P} \exp\left[i\int_{Z(0)}^{Z(T)} (P dX-XdP) \hat{I} + dX \hat{Y} - dP \hat{X}\right]
\end{align}
where $\hat{I}, \hat{X}, \hat{Y}, \hat{Z}$ are the Pauli matrices defined within instantaneous collision state subspace, see Appendix~\ref{appdix:col_holonomy} for more details. It is evident that the path-dependent term $(P dX-XdP)$ corresonding to $\hat{I}$ merely induce a global phase. In contrast, the terms associated with $\hat{X}$ and $\hat{Y}$ are path-independent, determined solely by the relative positions of the starting and ending points. Consequently, the movement of the collision state along the closed path will only result in inconsequential global phases. Therefore, operations related to $\hat{X}$ and $\hat{Y}$ in the instantaneous eigenstate space are produced exclusively along a open path. To validate it, we simulate the evolution of Hamiltonian with $\alpha_0(t)=\alpha_1(t)=\alpha(t)$ in Eq.~(\ref{eq:ham_eigen}). Without loss generity, we set the initial state to $\ket{\mathscr{C}}=c_0\ket{Z_0}+c_1\ket{Z_0,1}$ with $c_0=1.3$, $c_1=2e^{i\pi/6}$, the corresponding Wigner funciton is shown in Fig.~\ref{fig:col_move}(a), which starts from $Z_0=2+2i$ (blue dot) and evolves adiabatically to $Z_1=-1.7-2.3i$ (purple dot) along a blue-purple gradient path.  Fig.~\ref{fig:col_move}(b) shows the Wigner function of the final state. Finally, the fidelity of the simulation to the ideal state given by Eq.~(\ref{eq:col_holonomy}) is 0.999 with the evolution time of $T=100/K$. This demonstrates the effectiveness of the adiabatic evolution in realizing the desired quantum holonomy of collision states, laying the foundation for further exploration and utilization in quantum applications.

\section{Preparation of arbitrary cat states}
\label{sec:preparation}
\begin{figure*}
    \centering
    \includegraphics[width=\linewidth]{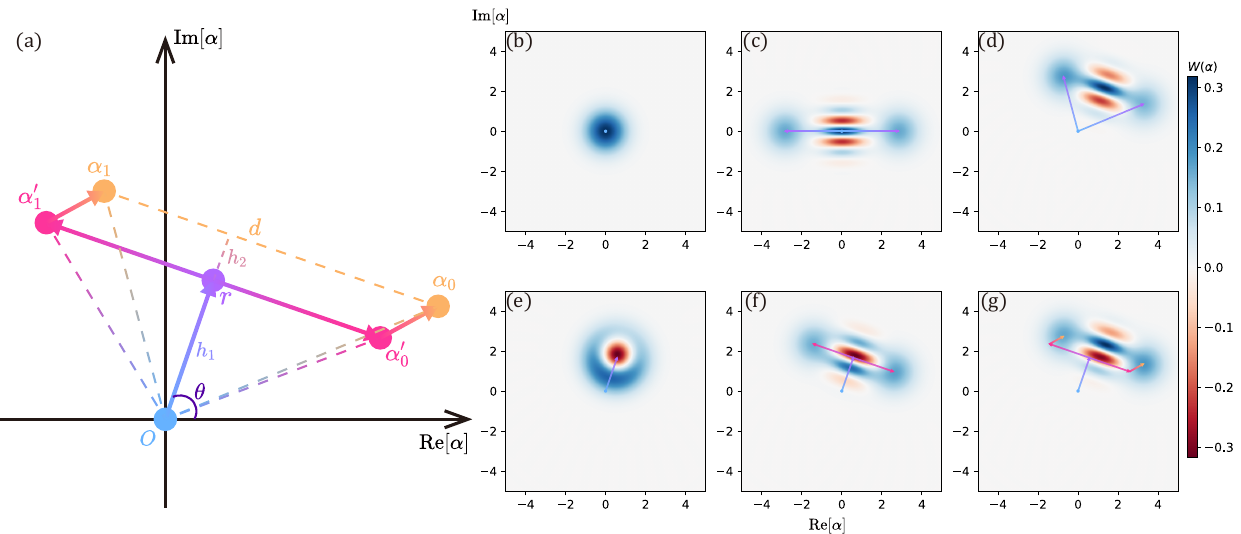}
    \caption{Cat state preparation. (a) Diagram of a holonomy-free preparation path for ACS. The Wigner function of the vacuum state (b) is the initial state. (c) The Wigner function of the final state of the traditional cat state, and the blue-purple gradient line represents the preparation path. (d) The Wigner function of the final state generated along the blue-purple gradient line for the direct preparation scheme of ACS. (e-g) The Wigner function at each stage of the phase-free preparation scheme shown in (a).}
    \label{fig:preparation}
\end{figure*}
In the preceding sections, we constructed a system in which ACS is the stable ground state. However, the system is not initially a cat state, and in practical experiments often the vacuum state. In this section, we demonstrate the procedure for preparing ACS from the vacuum state. We aim to prepare the target state $\mathcal{N}(\ket{\alpha_0}+\ket{\alpha_1})$, without loss of generality, where both components of the superposition state carry equal weight. Before delving into the preparation scheme of ACS, let us review its counterpart, the SCS, i.e. the case of $\alpha_0=-\alpha_1$. In this scenario, the cat state can be generated from the vacuum state by gradually increasing the amplitude of the two-photon driving to achieve the desired cat size. Fig.~\ref{fig:preparation}(c) shows the Wigner function of SCS preparation. This process can be explained intuitively using semiclassical metapotential described in Sec.~\ref{sec:metapotential}. Starting from the vacuum state in Fig.~\ref{fig:preparation}(b), the single potential well splits into two and evolves along the path of the blue-purple gradient line towards the purple point, ultimately yielding the SCS. To achieve an arbitrarily chosen ACS, drawing inspiration from the method used to prepare the SCS, it is intuitive to adopt a similar approach of gradually increasing $\alpha_0$ and $\alpha_1$ adiabatically to induce the potential wells to move along a straight path towards the target position. Fig.~\ref{fig:preparation}(d) shows the final state obtained by preparing the ACS along a straight path (blue-purple gradient line) starting from the vacuum state of Fig.~\ref{fig:preparation}(b).  It is evident that the final state has different weights and relative phases for $\ket{\alpha_0}$ and $\ket{\alpha_1}$. According to Eq.~(\ref{eq:geo_phase_coh}), the geometric phase of the coherent state along this preparation path should be zero. However, the final state is not perfect. This imperfection arises because, in the early stages of preparation, $\alpha_0$ and $\alpha_1$ are relatively close, effectively forming a near-collision state and introducing quantum holonomy to the instantaneous state. Unfortunately, aside from the complete collision state, we currently lack a precise method to predict the quantum holonomy of the near-collision state. Consequently, the adverse effects of this direct preparation scheme cannot be easily mitigated.
\par To overcome this challenge, we propose a holonomy-free ACS preparation scheme relying on precise predictions of the quantum holonomy in both coherent and collision states. Our proposed scheme is illustrated in Fig.~\ref{fig:preparation}(a), involves adjusting the parameters of the Hamiltonian Eq.~(\ref{eq:ham_eigen}) to steer the vacuum state along a predetermined trajectory, thus eliminating the detrimental impact from quantum holonomy. Below, we elaborate on the different stages of the preparation process in detail, which is mainly divided into three steps:\\
\begin{itemize}
    \item \emph{Step 1}: Fix $\alpha_0=\alpha_1$ to establish the system in a collision state. Initiate from the vacuum state and adiabatically move from the origin $O$ (blue dot) along the blue-purple gradient line to $r$ (purple dot). During this process, ensure that the path $Or$ is perpendicular to the line $\alpha_0 \alpha_1$ corresponding to the target state, consequently the angle $\theta$ between $Or$ and $X$-axis can be uniquely determined. Assuming the length of $Or$ is $h_1$, according to Eq.~(\ref{eq:col_holonomy}), the final state is $\ket{\psi_1}=\mathcal{N}[\cos(h_1)\ket{r,0}-\sin(h_1)e^{i\theta}\ket{r,1}]$ with $\mathcal{N}$ being the normalization factor. The Wigner function illustrating this state is presented in Fig.~\ref{fig:preparation}(e).
    \item \emph{Step 2}: The potential wells are separated from point $r$ and moved along the paths $r\alpha_0^\prime$ and $r\alpha_1^\prime$ simultaneously, at the same speed, until the separation distance reaches $|d|=|\alpha_0-\alpha_1|$. During this process, the path $\alpha_0^\prime \alpha_1^\prime$ aligns with the direction of $\alpha_0 \alpha_1$ of the target state. This can be analogized to preparing a cat state in a rotated coordinate system. The state $\ket{\psi_1}$ can be rewritten as $\ket{\psi_1^\prime}=\mathcal{N}[\cos(h_1)\ket{r,0}-i\sin(h_1)\ket{r,1}]$. where $\ket{r,0}$ will generate plus state $\mathcal{N}_+(\ket{\alpha_0^\prime}+\ket{\alpha_1^\prime})$, $\ket{r,1}$ will generate minus state $\mathcal{N}_-(\ket{\alpha_0^\prime}-\ket{\alpha_1^\prime})$. Additionally, this process yields geometric phases. Because the separation process on both sides is identical, the near-collision state at the early stage remains stationary thus not contributing to quantum holonomy. However, a geometric phase is induced in the coherent state. According to Eq.~(\ref{eq:geo_phase_coh}), this geometric phase is $\phi_1=-2S_{\triangle{O\alpha_0^\prime}\alpha_1^\prime}$. In summary, we obtain the final state  $\ket{\psi_2}=\mathcal{N}(\ket{\alpha_0^\prime}+e^{i(2h_1+\phi_1)}\ket{\alpha_1^\prime})$ whose Wigner function is depicted in Fig.~\ref{fig:preparation}(f).
    \item \emph{Step 3}: Translate the line $\alpha_0^\prime \alpha_1^\prime$ associated with $\ket{\psi_2}$ along the pink-orange gradient path to the line $\alpha_0\alpha_1$ to reach the target position. According to Eq.~(\ref{eq:geo_phase_coh}), the geometric phase generated in this process is $\phi_2=2(S_{\triangle O\alpha_0 \alpha_0^\prime}+S_{\triangle O\alpha_1 \alpha_1^\prime})$. Therefore, the final state can be represented as $\ket{\psi_3}=\mathcal{N}(\ket{\alpha_0}+e^{i(2h_1+\phi_1+\phi_2)}\ket{\alpha_1})$, whose Wigner function is shown in the Fig.~\ref{fig:preparation}(g).
\end{itemize}
Through the above process, the total phase becomes $\phi_{tol}=2h_1+\phi_1+\phi_2$. It is clear that $\phi_1+\phi_2=2(S_{\triangle O\alpha_0 \alpha_0^\prime}+S_{\triangle O\alpha_1 \alpha_1^\prime}-S_{\triangle{O\alpha_0^\prime}\alpha_1^\prime})=2(S_{\parallelogram \alpha_0 \alpha_1 \alpha_1^\prime \alpha_0^\prime}-S_{\triangle O\alpha_0 \alpha_1})=|d|(h_2-h_1)$. To achieve holonomy-free preparation, we should set the total phase $\phi_{tol}=2k\pi (k\in N)$ (with $(2k+1)\pi$ to prepare minus state), so the length of the \emph{Step 1} can be determined as
\begin{align}
    h_1 = \min \left\{ \left| \frac{H|d|-2k\pi}{2(|d|-1)} \right|, k \in N \right\}.
\end{align}
Where $H=h_1+h_2$ and separation distance $|d|=|\alpha_0-\alpha_1|$ is determined by known target parameters $\alpha_0,\alpha_1$. Utilizing the geometric principle that the hypotenuse of a triangle is longer than the other sides, minimizing $h_1$ensures the shortest total path. Once $h_1$ is determined, the entire trajectory is strictly defined. We employed this holonomy-free preparation scheme to generate an ACS with $\alpha_0=2.5e^{i\pi/8}, \alpha_1=2 e^{i7\pi/12}$. Fig.~\ref{fig:preparation} illustrates the preparation path (a) and the Wigner function (e-g) at each step. It is evident that the target state was successfully prepared with a fidelity of $0.999$.

\section{Holonomic universal quantum computing}
\label{sec:universal}
As described in Sec.~\ref{sec:stablize_loss}, a cat state with a large protected energy gap is highly resistant to resonator dephasing noise and ensures the stability of the manifold under photon loss. The quantum information encoded in the superposition coefficients is preserved despite photon loss and can be corrected by parity checks \cite{leghtas2013hardware, puri2019stabilized, cochrane1999macroscopically}. Consequently, this configuration is referred to as a cat qubit, offering a substantial advantage in noise robustness compared to traditional two-level systems.
\begin{figure}
    \centering
    \includegraphics[width=\linewidth]{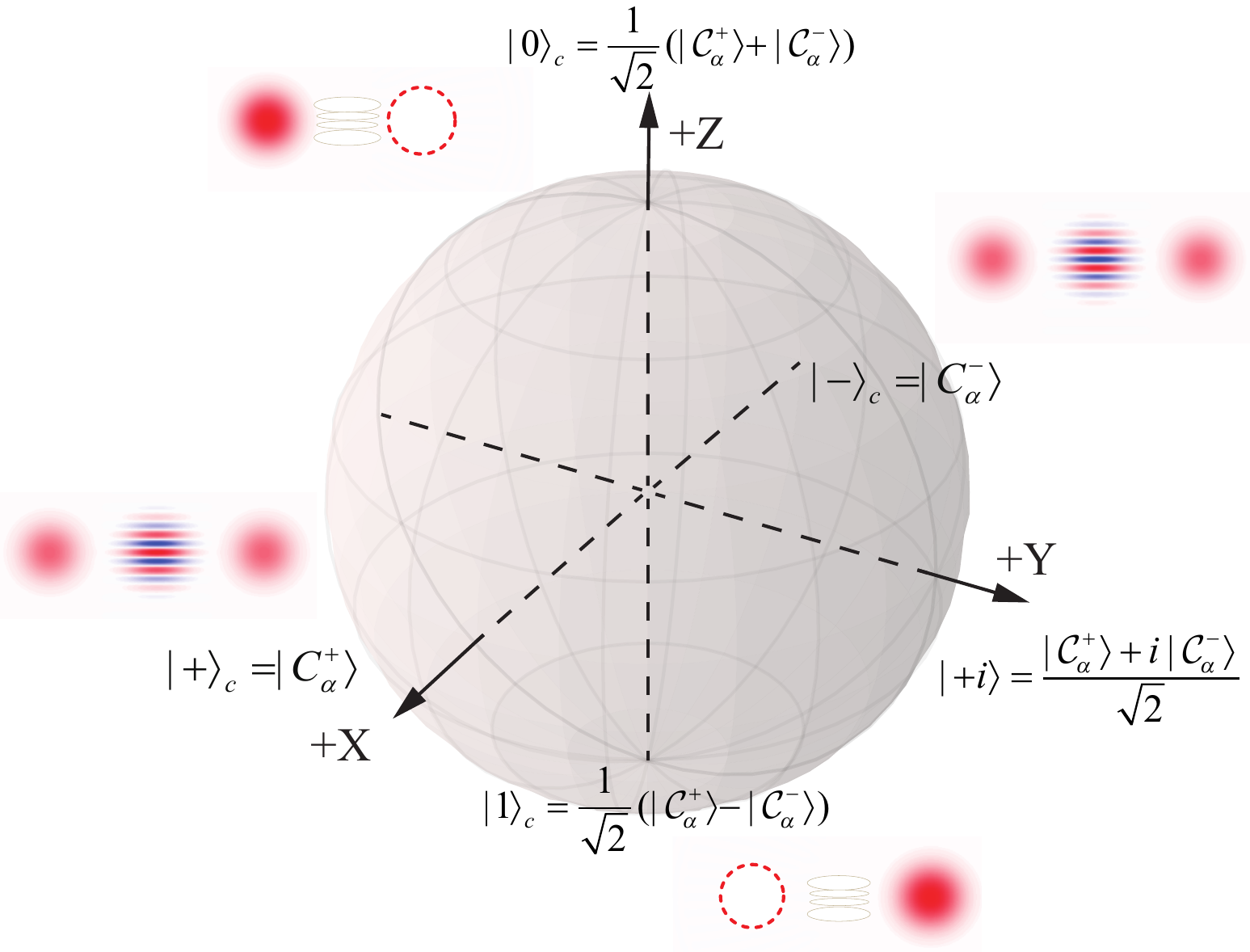}
    \caption{The Bloch sphere of cat qubit. The plus (minus) cat state $\ket{\mathcal{C}_\pm}$ is encoded as a logical qubit $\ket{0},\ket{1}$}.
    \label{fig:cat_bloch}
\end{figure}
As shown in the Fig.~\ref{fig:cat_bloch}, the Bloch sphere represents a cat qubit. The plus and minus cat states $\ket{\mathcal{C}_\pm}$, being perfectly orthogonal, are encoded as the logical qubit states $\ket{0}$ and $\ket{1}$. The X-axis is encoded as a superposition of the cat states, maintaining a strict orthogonal relationship and approximating coherent states $\ket{\pm \alpha}$ at large $\alpha$. This encoding forms the basis of a cat qubit. Furthermore, quantum holonomy, a global effect in parameter space, exhibits insensitivity to certain local noise. By leveraging these two advantages, we aim to construct universal holonomic gates for cat qubits in this section.
\subsection{Single-qubit gates}
\label{sec:single_qubit_gates}
Let us start with the single-qubit gate. Any state of cat qubit can be represented as a unit vector $\ket{\psi}=\cos \frac{\theta}{2} \ket{\mathcal{C}+} + \sin \frac{\theta}{2} e^{i\varphi} \ket{\mathcal{C}-}$ on the Bloch sphere, characterized by the angles $\theta$ and $\varphi$. The single-qubit gate controls the rotation of this state vector on the Bloch sphere. There are three common used rotations $R_x(\theta), R_y(\lambda), R_z(\varphi)$, defined as the exponentials of the corresponding Pauli matrices
\begin{align}
\label{eq:Rx}
&R_x(\theta)\equiv e^{-i\theta X/2} = \begin{pmatrix}
\cos \frac{\theta}{2} &-i\sin \frac{\theta}{2}\\
-i\sin \frac{\theta}{2} & \cos\frac{\theta}{2}
\end{pmatrix}\\
\label{eq;Ry}
&R_y(\lambda)\equiv e^{-i\lambda Y/2} = \begin{pmatrix}
\cos \frac{\lambda}{2} &-\sin \frac{\lambda}{2}\\
\sin \frac{\lambda}{2} & \cos\frac{\lambda}{2}
\end{pmatrix}\\
\label{eq:Rz}
&R_z(\varphi)\equiv e^{-i\varphi Z/2} = \begin{pmatrix}
e^{-i\varphi/2} &0\\
0 &e^{i\varphi/2}
\end{pmatrix}.
\end{align}
Holonomic $R_x(\theta)$ and $R_z(\varphi)$ gates were proposed by Albert et al. in 2016 \cite{albert2016holonomic}. Here, we will review these gates and, for the first time, demonstrate the implementation of the holonomic $R_y(\lambda)$ gate, thereby completing the set of Pauli rotations.
\begin{figure}
    \centering
    \includegraphics[width=\linewidth]{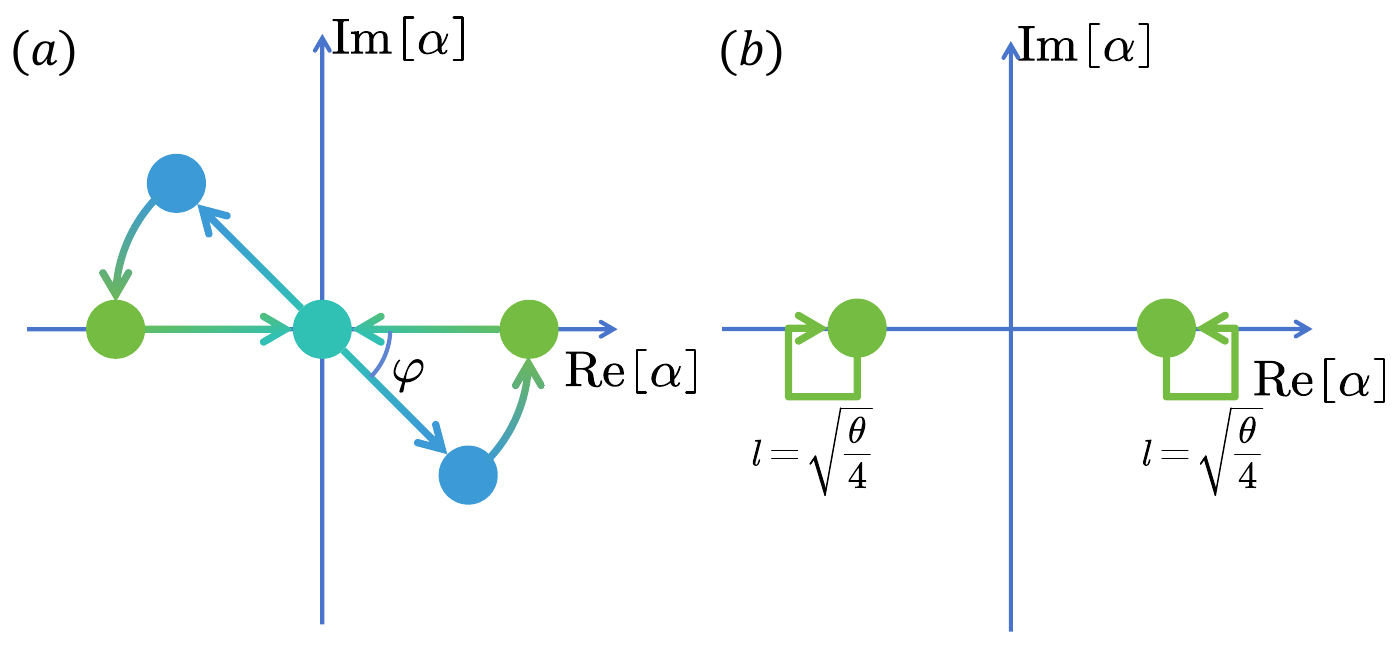}
    \caption{The trajectory to implement $R_z(\varphi)$ (a) and $R_x(\theta)$ (b) gate with cat qubit.}
    \label{fig:RxAndRz}
\end{figure}
\subsubsection{$R_z(\varphi)$ gate}
The trajectory to achieve $R_z(\varphi)$ gate is shown in Fig.~\ref{fig:RxAndRz}(a), without loss of generality, we illustrate this process with an arbitrary initial state $\ket{\psi_0}=c_0\ket{\mathcal{C}_+}+c_1\ket{\mathcal{C}_-}$. The implementation can be divided into three steps:
\begin{itemize}
    \item \emph{Step 1}: Adiabatically move the potential wells from $\alpha$ and $-\alpha$ at the same speed until they collide at the origin. The final state of this step is $\ket{\psi_1}=c_0\ket{0}+c_1\ket{1}$.
    \item \emph{Step 2}: Separate the potential wells along a line at an angle of $-\varphi$ from the X-axis until they reach the positions $\alpha e^{-i\varphi}$ and $-\alpha e^{-i\varphi}$. This process is similar to \emph{Step 2} for ACS preparation. In this rotated coordinate system, the state $\ket{\psi_1}$ can be rewritten as $\ket{\psi_1^\prime}=c_0\ket{0}+c_1e^{i\varphi}\ket{1}$. With $\ket{0}$ generating $\ket{\mathcal{C}_+^\prime}$ and $\ket{1}$ generating $\ket{\mathcal{C}_-^\prime}$, the final state of this step is $\ket{\psi_2}=c_0\ket{\mathcal{C}_+^\prime}+c_1e^{i\varphi}\ket{\mathcal{C}_-^\prime}$.
    \item \emph{Step 3}: Rotate counterclockwise by the angle $\varphi$ to return to the original cat code space to complete the operation. The final result is $\ket{\psi_3}=c_0\ket{\mathcal{C}+}+c_1e^{i\varphi}\ket{\mathcal{C}-}$.
\end{itemize}
From Eq.~(\ref{eq:Rz}), it is obvious that the above process implements the $R_z(\varphi)$ gate with an nonsignificant global phase $e^{-\varphi/2}$.

\subsubsection{$R_x(\theta)$ gate}
\par Next, we discuss the implementation of the $R_x(\theta)$ gate. Its trajectory is shown in Fig.~\ref{fig:RxAndRz}(b). We use the same arbitrary initial state $\ket{\psi_0}$ as mentioned for the $R_z$ gate. Let $\ket{\alpha}$ move counterclockwise and $\ket{-\alpha}$ move clockwise in phase space along a closed path with an area of $\theta/4$. According to Eq.~(\ref{eq:geo_phase_coh}), $\ket{\alpha}$ and $\ket{-\alpha}$ acquire phases of $-\theta/2$ and $\theta/2$ respectively. The final state of this process is: 
$\ket{\psi_1}=(\mathcal{N}_+ c_0 + \mathcal{N}_- c_1) e^{-i\theta/2}\ket{\alpha} + (\mathcal{N}_+ c_0 - \mathcal{N}_- c_1) e^{i\theta/2}\ket{-\alpha} = \cos(\theta/2)\ket{\mathcal{C}_+}+\sin(\theta/2)\ket{\mathcal{C}_-}$.
Thus, we achieve the $R_x(\theta)$ gate. To standardize the path, we can confine the trajectory to a square. Consequently, the side length can be determined as $l=\sqrt{\theta/4}$.

\subsubsection{$R_y(\lambda)$ gate}
\begin{figure*}
    \centering
    \includegraphics[width=\linewidth]{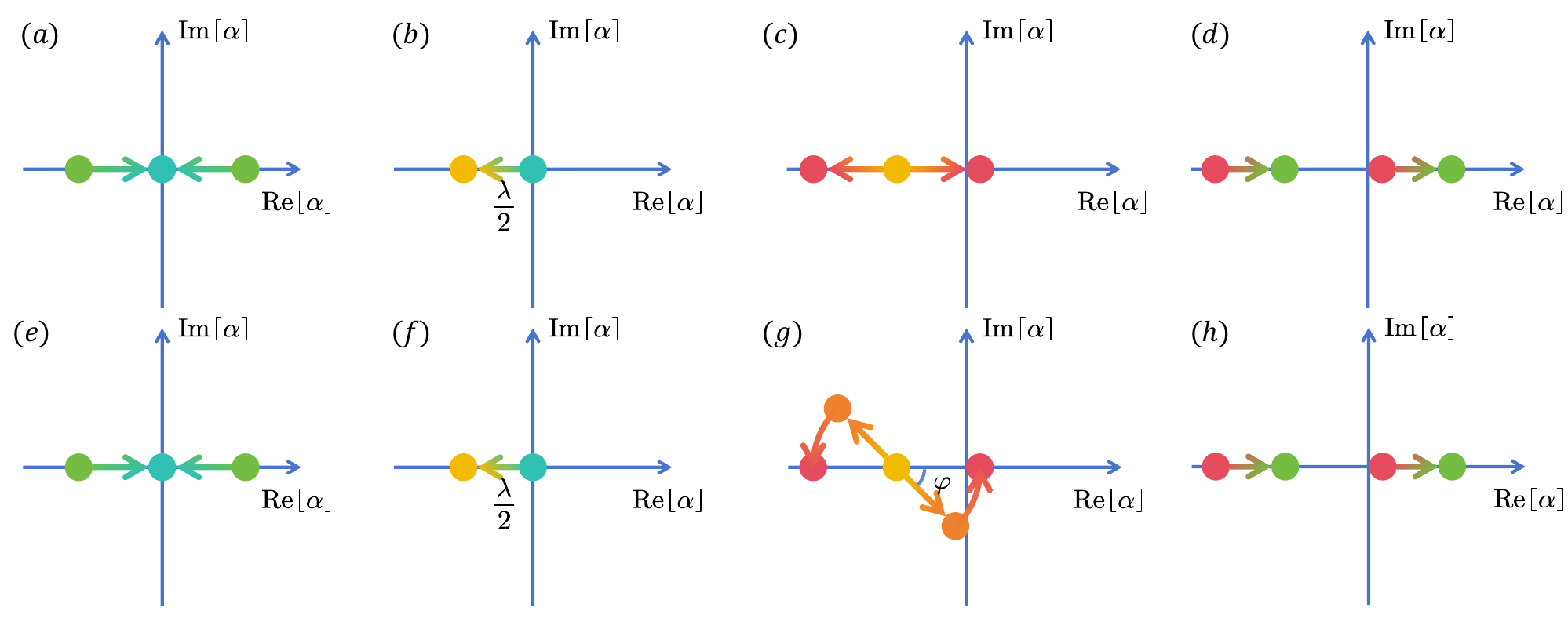}
    \caption{The trajectory to implement $Ry(\lambda)$ gate (a-d) and minimum-step $Rz(\varphi)R_y(\lambda)$ gate (e-h) with cat qubit.}
    \label{fig:RyAndRxRy}
\end{figure*}
\par In this section, we proposed the holonomic $R_y(\lambda)$ gate for the cat qubit for the first time. Although a universal single qubit operation can be constructed using the X-Z decomposition $U(\alpha, \beta, \gamma)=R_z(\alpha)R_x(\beta)R_z(\gamma)$, having an independent implementation of the $R_y$ gate can significantly enhance the flexibility of quantum circuits. In some cases, it reduces the cost associated with unnecessary combining $R_x$ and $R_z$ gates and decreases the circuit depth. Furthermore, for the important Hadamard gate, which transforms the basis, the X-Z decomposition requires three rotations, while the X-Y or Y-Z decomposition needs only two rotations. Thus, an independent $R_y$ gate offers substantial advantages. Next, we elaborate on the implementation of the $R_y$ gate, which utilizes the quantum holonomy of the collision state proposed in Sec.~\ref{sec:holonomy}. Fig.~\ref{fig:RyAndRxRy}(a-d) shows the evolution path of the holonomic $R_y(\lambda)$ gate, which can be divided into four steps:
\begin{itemize}
    \item \emph{Step 1}:  This step mirrors \emph{Step 1} of the $R_z$ gate, as shown in Fig.~\ref{fig:RyAndRxRy}(a), $\alpha$ and $-\alpha$ are adiabatically moved to the origin at the same speed, causing the two separated potential wells to collide and merge. Using the same arbitary initial state as used in $R_x$ gate, the final state of this step is $\ket{\psi_1}=c_0\ket{0}+c_1\ket{1}$.
    \item \emph{Step 2}: Move the collision potential well with a distance $-\lambda/2$ along the X axis as shown in Fig.~\ref{fig:RyAndRxRy}(b). According to the quantum holonomy of the collision state in Eq.~(\ref{eq:col_holonomy}), the final state of this step is $\ket{\psi_2}=e^{-i\lambda \hat{Y}/2} \ket{\psi_1^\prime} = R_y(\lambda) \ket{\psi_1^\prime}$.
    \item \emph{Step 3}: Separate the potential well along the X-axis until the distance is $2\alpha$ as shown in Fig.~\ref{fig:RyAndRxRy}(c). This is equivalent to a displaced version of the cat state preparation, resulting in an intermediate cat state $\ket{\psi_3}= R_y(\lambda) (c_0\ket{\mathcal{C}_+^\prime}+c_1\ket{\mathcal{C}_-^\prime})$.
    \item \emph{Step 4}: Displace the intermediate cat state along the X-axis by a distance of $\lambda/2$ to return to the original cat code space, the final state will be $\ket{\psi_4}= R_y(\lambda) \ket{\psi_0}$.
\end{itemize}
Clearly, the final state corresponds to the transformation of the $R_y(\lambda)$ gate applied to the initial state $\ket{\psi_0}$. Since the movement is performed along the X-axis, no geometric phases are introduced. This process effectively maps the holonomy of the collision state onto the cat state.
\subsubsection{Minimum step $R_z(\varphi)R_y(\lambda)$}
When executing quantum circuits, it is common to use a series of rotation gates rather than just one. An intuitive question arises: how can we make quantum circuits perform faster? This can be addressed at the algorithmic level by reducing the number of gates to make quantum circuits shallower, and at the physical implementation level by designing faster quantum gates. For the holonomic quantum gates presented in our paper, a single rotation gate could be accelerated in a non-adiabatic manner. However, each rotation gate still requires completing all the steps in its path. To reduce physical costs, we can attempt to eliminate redundant steps. The $R_z$ and $R_y$ gate share many similarities, such the processes of potential wells collision and separation. Therefore, we can consider combining these steps to accelerate the execution of quantum circuits. This combined process is shown in Fig.~\ref{fig:RyAndRxRy}(e-h). Aside from \emph{Step 3}, the process mirrors that of the $R_y$ gate. In the first two steps, (e) and (f), we obtain a displaced collision state with a holonomy from a $R_y(\lambda)$ operation. Then, applying a displaced version of \emph{Step 2} of the $R_z(\varphi)$ gate, as shown in Fig.~\ref{fig:RyAndRxRy}(g), results in an intermediate state  $R_z(\varphi)R_y(\lambda)\ket{\psi_0^\prime}$ where $\ket{\psi_0^\prime}=D(-\lambda/2)\ket{\psi_0}$. Finally, as shown in Fig.~\ref{fig:RyAndRxRy}(h), displacing the intermediate cat state back to the original code space yields the final result $\ket{\psi_f}=R_z(\varphi)R_y(\lambda)\ket{\psi_0}$.
For the $R_z(\varphi)R_y(\lambda)$ combined gates, we implement this operation exactly in the miminum steps. When executing some quantum circuits, if there has a $R_y$ gate concatenated by a $R_z$ gate, we can use this minimum-step implementation to obtain an acceleration. For example, Hadamard gate, the most basic operation on which many quantum algorithms such as Shor algorithm and quantum Fourier transform (QFT) depend, can be Z-Y decomposed as $R_z(\pi)R_y(-\pi/2)$, which can be realized by using our scheme.
\subsection{controlled gates for mutiple qubits}
\label{sec:controlled_gate}
In the previous content, we built universal single-qubit gates, however, full universal quantum computing also requires multi-qubit gates, in this part, we will explore how to implement. Let us start with the two-qubit gate, the most common is the controlled gate, where one qubit acts as the control qubit and the other acts as the target qubit is subjected to a unitary operation depending on the state of the control qubit. Returning to our system, without generality, we encode the 0 and 1 of the qubit as $\alpha_0$ and $\alpha_1$. For the traditional cat code, just set $\alpha_0=\alpha$ and $\alpha_1=-\alpha$. To implement the two-qubit control gate with holonomical way, we propose the following Hamiltonian
\begin{align}
\label{eq:ham_control_2qubits}
\begin{split}
    \hat{H} &= (\hat{a}_0^\dagger - \alpha_0^*)(\hat{a}_0^\dagger - \alpha_1^*)(\hat{a}_0 - \alpha_0 )(\hat{a}_0 - \alpha_1 )\\
    &+ \left[\hat{a}_1^\dagger -  \alpha_2^{ *} \left(\frac{\alpha_0^*-\hat{a}^\dagger_0}{\alpha_0^*-\alpha_1^*}\right) -  \alpha_4^{*} \left(\frac{\alpha_1^*-\hat{a}^\dagger_0}{\alpha_1^*-\alpha_0^*}\right) \right]\\
    &* \left[\hat{a}_1^\dagger -  \alpha_3^{*} \left(\frac{\alpha_0^*-\hat{a}^\dagger_0}{\alpha_0^*-\alpha_1^*}\right) -  \alpha_5^{*} \left(\frac{\alpha_1^*-\hat{a}^\dagger_0}{\alpha_1^*-\alpha_0^*}\right)\right]\\
    &* \left[\hat{a}_1 -  \alpha_2 \left(\frac{\alpha_0 -\hat{a}_0}{\alpha_0-\alpha_1}\right) -  \alpha_4 \left(\frac{\alpha_1 -\hat{a}_0}{\alpha_1-\alpha_0}\right)\right]\\
    &* \left[\hat{a}_1-  \alpha_3 \left(\frac{\alpha_0 -\hat{a}_0}{\alpha_0-\alpha_1}\right) -  \alpha_5 \left(\frac{\alpha_1 -\hat{a}_0}{\alpha_1-\alpha_0}\right)\right],
\end{split}
\end{align}
where subscript 0 represents the control qubit $Q_0$, and subscript 1 represents the target qubit $Q_1$. Obviously, the above equation is a Hamiltonian dependent on the state of $Q_0$. Before the specific discussion, it needs to be clear that when doing a two-qubit control gate, $Q_0$ must be in a cat state, that is, the potential wells not merge, meeting the condition $\alpha_0 \neq \alpha_1$. Under this premise, when $Q_0$ is in $\ket{\alpha_0}$, the Hamiltonian of $Q_1$ can be written as
\begin{align}
\hat{H}_{\alpha_0}^{Q_1} &= \left(\hat{a}_1^\dagger  -  \alpha_4^{*} \right)\left(\hat{a}_1^\dagger -  \alpha_5^{*} \right) \left(\hat{a}_1 -  \alpha_4 \right)\left(\hat{a}_1 -  \alpha_5\right),
\end{align}
and it can be seen that the state of $Q_1$ is determined by $\alpha_4$ and $\alpha_5$. When $Q_0$ is in $\ket{\alpha_1}$, the Hamiltonian of $Q_1$ can be written as
\begin{align}
\hat{H}_{\alpha_1}^{Q_1} &= \left(\hat{a}_1^\dagger  -  \alpha_2^{*} \right)\left(\hat{a}_1^\dagger -  \alpha_3^{*} \right) \left(\hat{a}_1 - \alpha_2 \right)\left(\hat{a}_1 -  \alpha_3\right)
\end{align}
and the state of $Q_1$ is determined by $\alpha_2$and $\alpha_3$. So far, it can be seen that the Eq.~(\ref{eq:ham_control_2qubits}) can realize the two-qubit cat state $c_0\ket{\alpha_0, \alpha_4} + c_1\ket{\alpha_0, \alpha_5} + c_2 \ket{\alpha_1, \alpha_2} + c_3\ket{\alpha_1, \alpha_3}$, where the parameters $\alpha_0, \alpha_1,\alpha_2, \alpha_3, \alpha_4, \alpha_5$ can be adjusted arbitrarily. When limiting to a two-qubit control gate, we only need to set $\alpha_4=\alpha_0, \alpha_5=\alpha_1$ to realize that the state of $Q_1$ remains unchanged when $Q_0$ is in $\ket{\alpha_0}$. In addition, by adjusting $\alpha_2, \alpha_3$ with $\alpha_0$ and $\alpha_1$ as initial values, moving them along the path corresponding to the single-qubit operation in the phase space, which can apply a unitary operation to $Q_1$ when $Q_0$ is in $\ket{-\alpha}$. Thus, we achieve a $C-U$ gate where $U$ can be any operations proposed in the single-qubit gate in Sec.~\ref{sec:single_qubit_gates}. 
\par It is obvious that we can extend the above two-qubit control gate to multi-qubit systems. For a system has $n$ qubits, it is numbered from 0, where the first $m$ qubits are the control qubits and the next $n-m$ qubits are the target qubits. To implement such a multi-qubit control gate, we set the Hamiltonian of all control qubits to be the same as $Q_0$ in a two-qubit system
\begin{align}
\hat{H}^c_i = (\hat{a}_i^\dagger - \alpha_0^*)(\hat{a}_i^\dagger - \alpha_1^*)(\hat{a}_i - \alpha_0 )(\hat{a}_i - \alpha_1 ),
\end{align}
where the subscript $i$ represents the $i$-th control qubit. The above Hamiltonian can put the control qubit in an arbitrary cat state $\mathcal{N}(\ket{\alpha_0}+\ket{\alpha_1})$. In addition, all target qubit Hamiltonians should also be similar to $Q_1$ in a two-qubit system, as shown below
\begin{align}
\hat{H}_k^t &= \left[\hat{a}_k^\dagger -  \alpha_2^{*} \left(\frac{\prod_{i=0}^{m-1}\left(\alpha_0^*-\hat{a}^\dagger_i\right)}{\left(\alpha_0^*-\alpha_1^*\right)^m}\right) \right.\\
&\quad -\left.\alpha_4^{*} \left(\frac{\sum_{j=1}^{2^{m}-1}\prod_{i=0}^{m-1}\left(\alpha_{l(i,j)}^*-\hat{a}^\dagger_i\right)}{-\left(\alpha_0^*-\alpha_1^*\right)^m}\right) \right]\\
&\times \left[\hat{a}_k^\dagger -  \alpha_3^{*}\left(\frac{\prod_{i=0}^{m-1}\left(\alpha_0^*-\hat{a}^\dagger_i\right)}{\left(\alpha_0^*-\alpha_1^*\right)^m}\right) \right.\\
&\quad -\left.\alpha_5^{*} \left(\frac{\sum_{j=1}^{2^{m}-1}\prod_{i=0}^{m-1}\left(\alpha_{l(i,j)}^*-\hat{a}^\dagger_i\right)}{-\left(\alpha_0^*-\alpha_1^*\right)^m}\right)\right]\\
&\times \left[\hat{a}_k -  \alpha_2 \left(\frac{\prod_{i=0}^{m-1}\left(\alpha_0-\hat{a}^\dagger_i\right)}{\left(\alpha_0-\alpha_1\right)^m}\right) \right.\\
&\quad \left.-  \alpha_4 \left(\frac{\sum_{j=1}^{2^{m}-1}\prod_{i=0}^{m-1}\left(\alpha_{l(i,j)}-\hat{a}^\dagger_i\right)}{-\left(\alpha_0-\alpha_1\right)^m}\right) \right]\\
&\times \left[\hat{a}_k -  \alpha_3\left(\frac{\prod_{i=0}^{m-1}\left(\alpha_0-\hat{a}^\dagger_i\right)}{\left(\alpha_0-\alpha_1\right)^m}\right) \right.\\
&\quad \left.-  \alpha_5 \left(\frac{\sum_{j=1}^{2^{m}-1}\prod_{i=0}^{m-1}\left(\alpha_{l(i,j)}-\hat{a}^\dagger_i\right)}{-\left(\alpha_0-\alpha_1\right)^m}\right)\right],
\end{align}
where the subscript $k$ represents the $k$-th target qubit, and the subscript $l\in \{0,1\}$ is equal to the $i$-th value of the binary representation of $j$. This Hamiltonian can make the behavior of the target qubits depend on the control qubits. When the state of the control qubits is in $\otimes_{i=0}^{m-1} \ket{\alpha_1}$, the state of all target qubits is determined by $\alpha_2, \alpha_3$, otherwise, the state of the all target qubits is determined by $\alpha_4,\alpha_5$. Therefore, the total Hamiltonian of the multi-qubit control gate can be expressed as $\hat{H} = \sum_{i=0}^{m-1}\hat{H}_i^c + \sum_{k=m}^{n-1} \hat{H}_k^t$. At this point, we can implement a multi-qubit gate with any control qubits and any target qubits. For example, we can set $n=3$ and $m=2$ to realize a three-qubit Toffoli gate.
\section{Physical realization in superconducting circuits}
\label{sec:phys_realize}
\begin{figure}
    \centering
    \includegraphics[width=\linewidth]{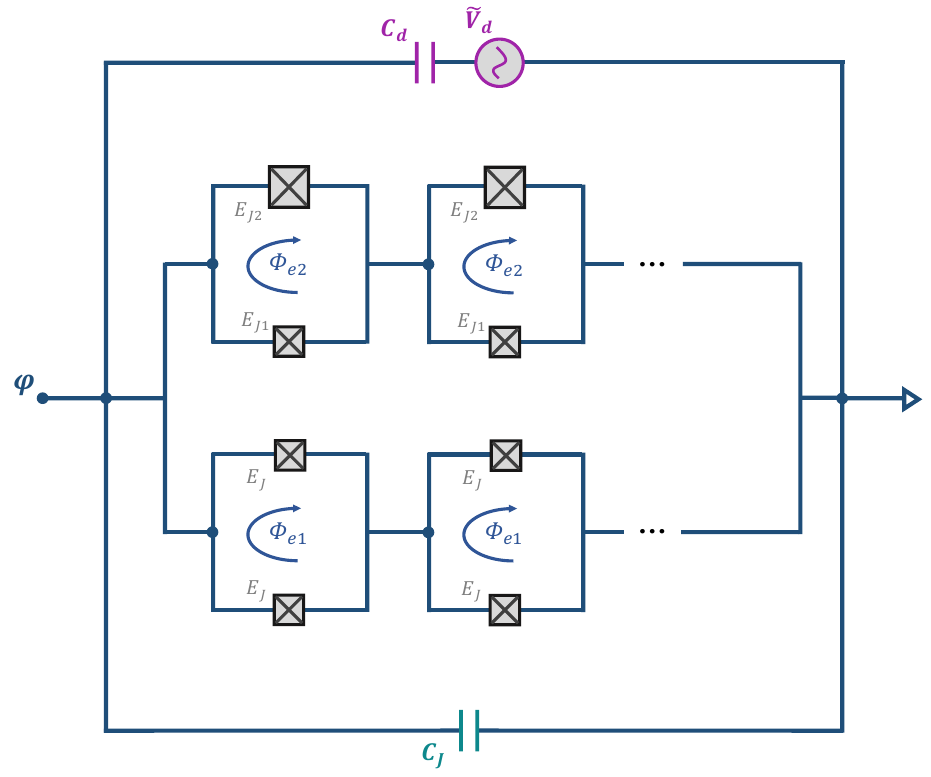}
    \caption{The physical realization circuit corresponding to the Hamiltonian of Eq.~(\ref{eq:ham_KNR}) consists of an intrinsic capacitor, a symmetric SQUID array, an asymmetric SQUID array and a capacitive coupled voltage source in parallel.}
    \label{fig:physical_circuit}
\end{figure}

Different from the traditional cat state, the parametric driving Hamiltonian we proposed in Eq.~(\ref{eq:ham_KNR}) not only includes nonlinear terms and second-order driving, but also involves detuning, first-order driving and third-order driving. The first and second order drivings can be realized by using capacitive coupled excitation source and flux-driven symmetric SQUID respectively, while third order drives require additional flux-driven asymmetric SQUID. Combining these knowledges, we propose the physical circuit shown in Fig.~\ref{fig:physical_circuit}, which consists of an intrinsic capacitor, a capacitive coupled voltage source, a symmetric SQUID array and an asymmetric SQUID array, noting that the purpose of using an N-dimensional array is to reduce nonlinearity, making it easier to achieve larger cat size.
The Hamiltonian of the above circuit is
\begin{align}
\begin{split}
    \hat H &= 4E_C \hat n^2 -2NE_J|\cos \varphi_{e1}| \cos (\hat{\varphi}/N) \\
    &-NE_{J_+}|\cos \varphi_{e2}|\cos (\hat{\varphi}/N)\\
    &-NE_{J_-}\sin \varphi_{e2}\sin (\hat{\varphi}/N) + \frac{4E_C C_d \tilde{V}_d}{e} \hat n,
\end{split}
\end{align}
where $\varphi_{e_i} = \pi \Phi_{ei} / \Phi_0$, $\Phi_0$ is the flux quanta, and $E_{J_\pm}=E_{J_1}\pm E_{J_2}$ represents the sum and difference of the Josephson inductive energies of the asymmetric SQUID. According to reference \cite{wang2019quantum}, the flux drivings in this Hamiltonian are a tiny oscillation around the mean value, it can be truncated to the first order to get $E_J + \delta_1 E_J\cos(\omega_2 t+\phi_2), E_{J_\pm} + \delta_2 E_{J_\pm}\cos(\omega_3 t+\phi_3)$. In addition, by setting the driving of the voltage source to $\tilde{V}_d=V_d\cos(\omega_1 t + \phi_1)$, we can get the driving Hamiltonian
\begin{align}
\hat H &= 4E_C \hat n^2   +  \frac{4E_C C_d V_d}{e}\cos(\omega_1 t + \phi_1) \hat n\\
& - N\tilde{E}_J\cos (\hat{\varphi}/N)- NE_{J_-}\sin (\hat{\varphi}/N) \\
& -2\delta_1 N E_J \cos (\omega_2 t + \phi_2) \cos (\hat{\varphi}/N) \\
&- \delta_2 N E_{J_+} \cos(\omega_3 t + \phi_3) \cos (\hat{\varphi}/N) \\
& - \delta_2 N E_{J_-} \cos(\omega_3 t + \phi_3) \sin (\hat{\varphi}/N),
\end{align}
where $\tilde{E}_J=(2E_J + E_{J_+})$. Performing the Taylor expansion of $\cos (\hat{\varphi}/N), \sin (\hat{\varphi}/N)$ and truncating it to the fourth order, then applying the second quantization $\hat{n}=-in_0(\hat{a}-\hat{a}^\dagger), \hat{\varphi}=\varphi_0(\hat{a}+\hat{a}^\dagger)$ to the above equation, where $n_0=[\tilde{E}_J/(32NE_c)]^{\frac{1}{4}}, \varphi_0=[2NE_C/\tilde{E}_J]^{\frac{1}{4}}$. For simplicity, let $\hbar=1$ and ignore the constant term, we get the following Hamiltonian
\begin{align}
\hat H &= \omega_c \hat a^\dagger \hat a - \frac{E_C}{12N^2}(\hat a + \hat a^\dagger)^4 \\
& -NE_{J-}\left[ \left(\frac{\varphi_0}{N}\right)(\hat a + \hat a^\dagger) - \left(\frac{\varphi_0}{N}\right)^3\frac{(\hat a + \hat a^\dagger)^3}{3!} \right] \\
& -2\delta_1 NE_J \left[- \left(\frac{\varphi_0}{N}\right)^2\frac{(\hat a + \hat a^\dagger)^2}{2!} \right.\\
&\quad \left.+ \left(\frac{\varphi_0}{N}\right)^4\frac{(\hat a + \hat a^\dagger)^4}{4!}  \right] \cos (\omega_2 t + \phi_2) \\
& - \delta_2 NE_{J-} \left[ \left(\frac{\varphi_0}{N}\right)(\hat a+ \hat a^\dagger)\right. \\
&\quad \left.- \left(\frac{\varphi_0}{N}\right)^3 \frac{(\hat a+ \hat a^\dagger)^3}{3!} \right] \cos(\omega_3 t + \phi_3)\\
& - \delta_2 NE_{J+} \left[-\left(\frac{\varphi_0}{N}\right)^2 \frac{(\hat a+ \hat a^\dagger)^2}{2!} \right.\\
&\quad \left.+ \left(\frac{\varphi_0}{N}\right)^4 \frac{(\hat a + \hat a^\dagger)^4}{4!} \right] \cos(\omega_3 t + \phi_3)\\
& -i n_0\frac{4E_C C_d V_d}{e}\cos(\omega_1 t + \phi_1)  (\hat a - \hat a^\dagger),
\end{align}
where $\omega_c = \sqrt{8\tilde{E}_{J}E_C/N}$. Next, we set each driving frequencies as $\omega_1=\omega_2/2=\omega_3=\omega_p$, and transform the Hamiltonian of the above equation into a rotating frame with a frequency of $\omega_p$. Taking the ratation wave approximation to get the final Hamiltonian
\begin{align}
\hat H' = \Delta \hat a^\dagger \hat a + K \hat a^{\dagger 2} \hat a^2 +\left(\beta^*\hat a^{\dagger 2}+ \beta\hat a^2\right)\\
+\left ( \varepsilon^* \hat a^\dagger + \varepsilon \hat a \right) +\left(\eta^* \hat a^{\dagger 2}\hat a + \eta \hat a^\dagger \hat a^2 \right).
\end{align}
In the above equation, the Kerr nonlinearity is $K=-E_c/2N^2$, the detuning between the eigenfrequency and the rotating frequency is $\Delta=\omega_c + K/2 - \omega_p$, the different order driving strengths are 
\begin{align}
    &\beta=\frac{\delta_1 E_J \omega_c}{4\tilde{E}_J}e^{i\phi_2}\\
    &\eta=\frac{\delta_2 E_{J_-}\omega_c^{3/2}}{4\sqrt{N}\left(2\tilde{E}_J\right)^{3/2}}e^{i\phi_3}\\
    &\varepsilon=\epsilon+\xi,
\end{align}
the parameters in first order driving are defined as 
\begin{align}
    &\xi=-\delta_2E_{J_-}\sqrt{\frac{N\omega_c}{8\tilde{E}_J}}e^{i\phi_3}\\
    &\epsilon = -i n_0\frac{2E_C C_d V_d}{e} e^{i\phi_1}.
\end{align}
So far, we can use the physical circuit in Fig.~\ref{fig:physical_circuit} to implement the desired Hamiltonian in Eq.~(\ref{eq:ham_KNR}).
\section{Conclusion}
\label{sec:conclusion}
In this paper, we proposed a physical system to stabilize arbitrary coherent superposition cat states and provide a precise preparation method. Our system allows greater freedom in controlling cat states, enabling independent manipulation of the superimposed components. We visualize this control using semi-classical potential wells and study the collision of superimposed components, introducing the concept of quantum holonomy for collision states for the first time.
Additionally, we present universal quantum holonomic gates for cat qubits, including the first implementation method for the $R_y$ gate, thereby completing all Pauli rotation operations. We also provide the Hamiltonian for multi-qubit gates, enabling the realization of C-U gates with any nnn control and mmm target qubits, demonstrating significant practical value.
Finally, we demonstrate the realization of these theoretical results in superconducting circuits. Based on existing experimental technology, all operations can be implemented. Our study enhances the practicality of holonomic gates based on cat qubits and provides a platform with higher degrees of control to explore more interesting physics.

\begin{acknowledgments}
We would like to thank X. Wang for their valuable discussion. H. R. Li is supported by the National Natural Science Foundation of China (Grant No.11774284).
\end{acknowledgments}

\appendix
\section{Orthonormality of ACS}
\label{appdix:orth_norm}
An arbitrarily uniformly superposed ACS can be represented as $\ket{\mathcal{C}_\pm}=\mathcal{N}_{\pm}(\ket{\alpha_0} \pm \ket{\alpha_1})$. The orthogonality of any coherent state is
\begin{align}
    \braket{\alpha_0 | \alpha_1}&=\exp \left[-\frac{1}{2}\left(|\alpha_0|^2 + |\alpha_1|^2-2\alpha_0^* \alpha_1 \right) \right]\\
    &=\exp \left[-\frac{1}{2} \left(|d|^2 + 2i \rm{Im}[\alpha_0 \alpha_1^*] \right) \right]
\end{align}
Thus, the norm of ACS is
\begin{align}
    \braket{\mathcal{C}_+|\mathcal{C}_+} = \mathcal{N}_+^2 \left(2+2e^{-\frac{1}{2}|d|^2} \cos \gamma \right)\\
    \braket{\mathcal{C}_-|\mathcal{C}_-} = \mathcal{N}_-^2 \left(2-2e^{-\frac{1}{2}|d|^2} \cos \gamma \right)
\end{align}
where $\gamma = \rm{Im}[\alpha_0 \alpha_1^*]=|\alpha_0\alpha_1| \sin(\phi_0-\phi_1)$, with $\phi_i$ is the phase factor of complex value $\alpha_i$. Therefore, the normalization coefficient of ACS is
\begin{align}
    \mathcal{N}_\pm = \frac{1}{\sqrt{2(1\pm e^{-\frac{1}{2}|d|^2} \cos \gamma)}}.
\end{align}
Next, we explore its orthogonality,
\begin{align}
    \braket{\mathcal{C}_+|\mathcal{C}_-} = i \frac{e^{-\frac{1}{2}|d|^2} \sin \gamma}{\sqrt{1-e^{-|d|^2} \cos^2 \gamma}}.
\end{align}
Obviously, ACS is perfectly orthogonal when the phase difference is $\phi_0-\phi_1=\pi$.

\section{Quantum Holonomy of Collision state}
\label{appdix:col_holonomy}
As mentioned in the Sec.~\ref{sec:collision_states}, when the potential well collides at $\alpha$, the ground degenerate states is collision states $\{\ket{\alpha,0}, \ket{\alpha,1} \}$, spanning a 2-dimensional degenerate subspace. The collision states motion in its parameter space, i.e., phase space, produces quantum holonomy in Eq.~(\ref{eq:holonomy_def}), its generator $\Gamma$ can be represented as
\begin{align}
\label{eq:generator_ori}
    \Gamma(\mathbf{Z}) = i\begin{pmatrix}
        \bra{\mathbf{Z},0} \nabla \ket{\mathbf{Z},0} &\bra{\mathbf{Z},0} \nabla \ket{\mathbf{Z},1}\\
        \bra{\mathbf{Z},1} \nabla \ket{\mathbf{Z},0} &\bra{\mathbf{Z},1} \nabla \ket{\mathbf{Z}e,1}
    \end{pmatrix}
\end{align}
In the above equation, $\mathbf{Z}=X+iP$ is the position in phase space, and the space differential is $\nabla=\frac{\partial}{\partial X} + \frac{\partial}{\partial P}$. The displacement transformation of the differential operator gives
\begin{align}
\label{eq:nabla_displace}
    D^\dagger(\mathbf{Z}) \nabla D(\mathbf{Z}) = -iP+\hat{a}^\dagger - \hat{a} + iX + i(\hat{a}^\dagger + \hat{a}).
\end{align}
Combining Eq.~(\ref{eq:generator_ori}) and Eq.~(\ref{eq:nabla_displace}), we can get
\begin{align}
\label{eq:generator_final}
    \Gamma(\mathbf{Z})=\begin{pmatrix}
        PdX-XdP &-idX-dP\\
        idX-dP &PdX-XdP,
    \end{pmatrix}
\end{align}
in the Pauli space of the instantaneous collision state, it can be written as $\Gamma(\mathbf{Z})=(PdX-XdP)\hat{I} + dX \hat{Y} - dP \hat{X}$. Obviously, we can get the a unitary matrix in Eq.~(\ref{eq:col_holonomy}).
\nocite{*}

\bibliography{reference}

\end{document}